\documentclass{jpp}

\usepackage{graphicx}
\usepackage{natbib}
\usepackage[utf8]{inputenc}
\ifCUPmtlplainloaded \else
  \checkfont{eurm10}
  \iffontfound
    \IfFileExists{upmath.sty}
      {\typeout{^^JFound AMS Euler Roman fonts on the system,
                   using the 'upmath' package.^^J}%
       \usepackage{upmath}}
      {\typeout{^^JFound AMS Euler Roman fonts on the system, but you
                   dont seem to have the}%
       \typeout{'upmath' package installed. JPP.cls can take advantage
                 of these fonts, if you use 'upmath' package.^^J}%
      }
  \else
  \fi
\fi

\usepackage{bm}
\usepackage{color}
\usepackage{amssymb}
\usepackage{amsmath}
\newcommand{\z}{\mathbf{z}}

\newcommand{\vb}{\mathbf{v}}
\newcommand{\xb}{\mathbf{x}}
\newcommand{\D}[2]{\frac{\partial{#1}}{\partial{#2}}}

\ifCUPmtlplainloaded \else
  \checkfont{msam10}
  \iffontfound
    \IfFileExists{amssymb.sty}
      {\typeout{^^JFound AMS Symbol fonts on the system, using the
                'amssymb' package.^^J}%
       \usepackage{amssymb}%

      }{}
  \fi
\fi

\ifCUPmtlplainloaded \else
  \IfFileExists{amsbsy.sty}
    {\typeout{^^JFound the 'amsbsy' package on the system, using it.^^J}%
     \usepackage{amsbsy}}
    {}
\fi

\newsavebox{\astrutbox}
\sbox{\astrutbox}{\rule[-5pt]{0pt}{20pt}}

\newcommand\eero[1]{{\color{black}#1}}

\title[MC solution of minority particle Fokker-Planck equation]{Monte Carlo method and High Performance Computing for solving Fokker-Planck equation of minority plasma particles}

\author[E. Hirvijoki et al.]%
{E. Hirvijoki$^{1,2}$\thanks{email: eero.hirvijoki@chalmers.se}, T. Kurki-Suonio$^{2,1}$, 
S. Äkäslompolo$^2$, J. Varje$^2$, T. Koskela$^2$ and J. Miettunen$^2$%
}

\affiliation{
$^1$ Department of Applied Physics, Chalmers Univ. of Technology, Gothenburg, 41296, Sweden
\\
[\affilskip]
$^2$Department of Applied Physics, Aalto University, Espoo, 02015, Finland
}

\pubyear{2010}
\volume{650}
\pagerange{119--126}
\date{September 26 2014; revised February 05 2015; accepted February 06 2015.\footnote{doi:10.1017/S0022377815000203}}
\begin{document}

\maketitle

\begin{abstract}
This paper explains how to obtain the distribution function of minority ions in tokamak plasmas using the Monte Carlo method. Since the emphasis is on energetic ions, the guiding-center transformation is outlined, including also the transformation of the collision operator. Even within the guiding-center formalism, the fast particle simulations can still be very CPU intensive and, therefore, we introduce the reader also to the world of high-performance computing. The paper is concluded with a few examples where the presented method has been applied.
\end{abstract}


\section{Introduction}
The density of high-performance plasma is about $10^{20}$\,m$^{-3}$ and the volume of, for instance, the ITER vacuum vessel is over 800\,m$^3$. Therefore the total number of particles in the system is of the order of $10^{23}$. Furthermore, since the plasma consists of charged particles, the Coulomb interaction between them has, in principle, infinite range, and so the motion of each particle is connected to the motion of all other particles. Therefore, instead of trying to determine the dynamics of individual particles, a more efficient description for the plasma is needed. In fact, the behaviour of plasma at microscopic level is not even interesting in fusion applications: it is not important which particle is at a given position with a given velocity. Rather, it is the \emph{number} of particles in such a {\em phase-space} position that will determine the macroscopic behaviour of the plasma.

The {\em distribution function}, $f(\xb,\vb,t)$, gives the \emph{probability density} at time $t$ to find a particle in phase-space volume element  $d\xb\,d\vb$. There are two conventions for normalizing the distribution function: If $f$ is literally taken as a probability density, it is normalized to unity, $\int f d\xb\,d\vb = 1$  -- meaning that particle has to be found somewhere in the phase-space. If, on the other hand, $f$ is viewed as a distribution of the particles, then it is normalized according to $\int f(\xb,\vb)  d\xb\,d\vb = N$, meaning that when integrated over all velocities and across all the space, it has to give the total number of particles in the system. Regardless of the normalization, all the relevant macroscopic quantities can then be obtained as different velocity moments of the function $f$. For instance, the number density of particles $n(\mathbf r)$ is obtained by integrating over the velocity space (and possibly multiplying by $N$).

Adopting the distribution function, we have adopted a \emph{statistical} description of the plasma. For this to be a useful tool, we need a dynamic equation for the distribution function in place of the equations of motion for individual particles. In the vast majority of fusion problems, the equation to use is the so-called \emph{Fokker-Planck equation}. How to solve this equation is the topic of this paper.

The structure of the paper is as follows: In Sec.~\ref{sec:FP}, we present a derivation of the Fokker-Planck equation and in Sec.~\ref{sec:Langevin} relate it to stochastic processes, opening up the possibility of solving the Fokker-Planck equation by means of Monte Carlo methods. In Sec.~\ref{sec:FPcoefficients}, we give expressions for the coefficients in the Fokker-Planck equation for a few cases important for studying fast particle physics. Section~\ref{sec:GC} is devoted to eliminating the gyro motion that in many cases does not provide any useful information. The elimination is carried out using Lie-transformation perturbation methods. The details of the transformation are not only lengthy and tedious, but also require several clever tricks that are not obvious. As for anyone entering the field it is easy to become tangled in the details, losing the view of the process itself, we try to outline the process and give references to where to find those tedious details. Section~\ref{sec:HPC} is dedicated to high-performance computing, and in Sec.~\ref{sec:examples} we give a few examples of cases where the approach presented here has been found indispensable. 

\section{Fokker-Planck equation}\label{sec:FP}
Let us first take a look at a situation where there are no external electric or magnetic fields but, rather, the particles in the plasma move only in the presence of the microscopic fluctuating electromagnetic fields produced by other particles. \eero{These microscopic fields are assumed to be the ''left-overs'' as the true fields are averaged over some reasonable length scale to obtain the macroscopic fields, and the distribution function at macroscopic scales. In plasmas, this length scale is the \textit{Debye} length, and in typical fusion plasmas, the Debye length is small compared to the interesting macroscopic scales. As the interaction between the particles and these microscopic left-over fields essentially happens within a sphere with a radius of the size of the Debye length, the effects of these microscopic left-over fields are thus point-wise and referred to as the \textit{Coulomb collisions}. For an interested reader, the microscopic fluctuations are analyzed in detail in the book by~\citet{ichimaru1973basic}}.

We will also have to assume that the interparticle forces cause only small-angle deflections in particle trajectories. Since the Coulomb force has a long range, the assumption of only small-angle deflections is a reasonable in hot and dilute plasmas, such as most plasmas in tokamaks are: The deflection of particle's trajectory during a single collision event can be shown small. With these assumptions in mind, we are now in the position to illustrate a derivation of the Fokker-Planck equation.

\eero{Following~\citet{ichimaru1973basic}}, let us assume that at a time $t$ the particle is found at $\z = (\xb,\vb )$, and that at an instant later, it is found at $\z + \Delta$. The likelihood of the transition from $\z$ to $\z+\Delta$ during a (short) time $\tau$ is described by a transition probability $W_{\tau}(\z,\Delta)$, with normalization $\int W_{\tau}(\z,\Delta)d\Delta=1$. Then the probability density for finding the particle at $\z$ after time~$\tau$ is given by summing up all possible jumps that lead to $\z$
\begin{equation}
\label{eq:transition}
f(\z,t+\tau)=\int d\Delta\; f(\z-\Delta,t)W_{\tau}(\z-\Delta,\Delta).
\end{equation}
Assuming small-angle collisions, the steps $\Delta$ are small and one can carry out Taylor expansion of $f$ and $W_{\tau}$ around $\z$ 
\begin{multline}
\label{eq:expanded_fokker_planck}
f(\z,t+\tau)=\int d\Delta \left[ f(\z,t)W_\tau(\z,\Delta) - \D{}{\z}\left(\frac{}{}f(\z,t)W_\tau(\z,\Delta)\right)\cdot\Delta\right.
\\+\left.\frac{1}{2}\D{}{\z}\D{}{\z}\left(\frac{}{}f(\z,t)W_\tau(\z,\Delta)\right):\Delta\Delta+\mathcal{O}(\Delta\Delta\Delta)\right].
\end{multline}
Dividing by $\tau$ and rearranging the terms one obtains
\begin{multline}
\frac{f(\z,t+\tau)-f(\z,t)}{\tau}=-\D{}{\z}\cdot\left(f(\z,t)\frac{\langle\Delta\rangle}{\tau}\right)\\
+\frac{1}{2}\D{}{\z}\D{}{\z}:\left(f(\z,t)\frac{\langle\Delta\Delta\rangle}{\tau}\right)+\mathcal{O}\left(\frac{\langle\Delta\Delta\Delta\rangle}{\tau}\right),
\end{multline}
Here the expectation value with respect to the transition probability is defined as
$$\langle \dots \rangle=\int d\Delta W_{\tau}(\z,\Delta) \dots $$

When calculating the average change in the particle velocity due to Coulomb collisions, $\langle\Delta\vb\rangle$, and the second moment, $\langle\Delta\vb\Delta\vb\rangle$, the same logarithmic divergence appears as is found when calculating the Rutherford scattering cross section. When the appropriate cut-off for the integration is applied, i.e., limiting the range of Coulomb interaction within the Debye length, these first and second moments become proportional to the Coulomb logarithm, $\ln\Lambda$. On the contrary, higher order moments will not contain such divergencies. Therefore, if $\ln\Lambda\gg 1$, the first two moments $\frac{\langle\Delta\rangle}{\tau}$ and $\frac{\langle\Delta\Delta\rangle}{\tau}$ dominate higher order terms by the factor~$\ln\Lambda$. Typically, for fusion plasma, $\ln\Lambda \sim 15$, and it is justified to drop the higher order terms and write 
\begin{align}
\D{}{ t}f(\z,t)=-\D{}{\z}\cdot\left[(\mathbf{a}(\z,t))f(\z,t)\right]+\D{}{\z}\D{}{\z}:\left[\mathbf{D}(\z,t)f(\z,t)\right],
\end{align}
where the collisional friction (or drag) vector and diffusion tensor (the Fokker-Planck coefficients) are defined as
\begin{align} 
\mathbf{a}(\z,t)&=\lim_{\tau\rightarrow 0}\frac{\langle\Delta\rangle}{\tau},\\
\mathbf{D}(\z,t)&=\lim_{\tau\rightarrow 0}\frac{\langle\Delta\Delta\rangle}{2\tau}.
\end{align}

\begin{figure}
\centering
\includegraphics[width=0.5\textwidth]{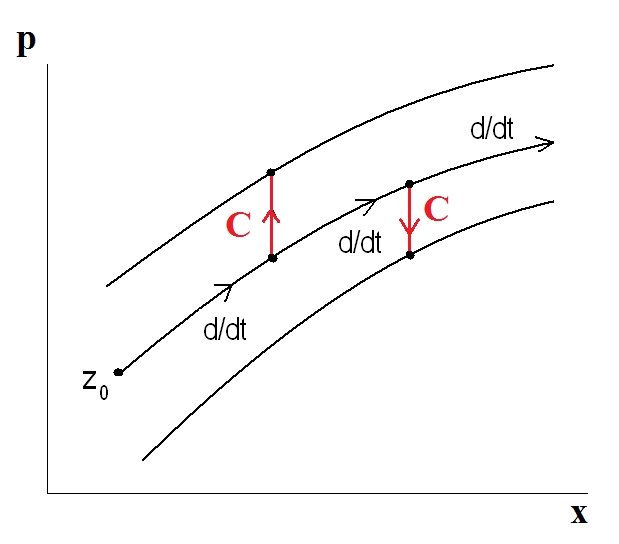}
\caption{\label{fig:Alain1}The effect of collisions in particle space. The solid lines represent the deterministic Hamiltonian trajectories in the phase-space and the red vertical arrows the effect of collisions. \emph{Courtesy of Alain Brizard}}
\end{figure}

\eero{To illustrate how the effect of macroscopic electric and magnetic fields can be included}, let us neglect the microscopic interparticle forces for a while and consider the transition probability to be a $\delta$-function peaked at the change in $\z$ due to macroscopic forces, i.e., $W_{\tau}=\delta(\Delta\z-\Delta)$. All the moments thus become $\langle(\Delta\z)^n\rangle=(\Delta\z)^n$. If $\tau$ is short, we find $\langle(\Delta\z)^n\rangle\approx(\dot{\z}\tau)^n$. In the limit  $\tau\rightarrow 0$ the result becomes exact, $\Delta\z\rightarrow\dot{\z}\tau$, and we obtain an exact result %
\begin{equation}
\D{}{t}f(\z,t)+\D{}{\z}\cdot\left(\dot{\z}f(\z,t)\right)=0,
\end{equation}
describing the time evolution of a function $f(\z(t),t)$ due to \eero{macroscopic forces}. This is the result one would expect, if only, e.g., Hamiltonian changes in the phase-space coordinates were allowed. 
Now the full equation for the distribution function, applicable to fusion plasmas, becomes
\begin{align}
\D{}{ t}f(\z,t)=-\D{}{\z}\cdot\left[(\dot{\z}+\mathbf{a}(\z,t))f(\z,t)\right]+\D{}{\z}\D{}{\z}:\left[\mathbf{D}(\z,t)f(\z,t)\right],
\end{align}
and it is often referred to as the \textit{kinetic equation} or the \textit{Fokker-Planck equation}. Figure~\ref{fig:Alain1} illustrates the role of different terms in the Fokker-Planck equation for a 1-dimensinal situation: the collisional terms kick the particles from one macroscopic trajectory to another. As was discussed earlier, that the Coulomb interaction due to microscopic fields is point-wise, the collisional kicks in the picture are necessarily vertical: \emph{the changes due to Coulomb collisions operate only in the particle's velocity space coordinates}. %

It is important to keep in mind that $\ln\Lambda$ determines the {\em validity} of the Fokker-Planck theory. The theory completely breaks down when $\ln\Lambda\rightarrow 1$ as the error is proportional to $1/\ln\Lambda$. \eero{Such situations are encountered, for instance, in dense and cold plasmas in white dwarfs, and in plasmas that are induced by laser pulses shot to solid targets. Such systems are outside to scope of Fokker-Planck theory}.

The collisional terms, $\mathbf{a}$ and $\mathbf{D}$, are often written separately from the Hamiltonian contribution:
\begin{align}
\D{}{ t}f+\D{}{\z}\cdot\left(\dot{\z}f\right)=\;C\left[f\right],
\end{align}
where the right-hand-side, $C\left[f\right]$, is called the \textit{collision operator}
\begin{align}
\label{eq:general_collision_operator}
C\left[f\right]=&-\D{}{\z}\cdot\left[\mathbf{a}f-\D{}{\z}\cdot\left(\mathbf{D}f\right)\right]\equiv -\D{}{\z}\cdot\mathbf{J},
\end{align}
and $\mathbf{J}$ is the collisional flux density. Notice that the divergence form of the collision operator guarantees conservation of particles: when integrating over velocity space, which yields the number of particles from the operator, it gives zero, i.e., no particles are created/destroyed in the collisions.

\section{Connection to stochastic processes: the Langevin equation}
\label{sec:Langevin}
In principle, the Fokker-Planck equation gives us a perfectly satisfactory description of fusion plasmas. In practice, however, it is impossible to solve directly in full six dimensions of the phase-space. Fortunately, already in the 1930's it was shown (by Kolmogorov and others) that if the distribution function $f$ is the solution of the Fokker-Planck equation
\begin{equation*}
\D{}{ t}f(\z,t)=-\D{}{\z}\cdot\left[(\dot{\z}+\mathbf{a}(\z,t))f(\z,t)\right]+\D{}{\z}\D{}{\z}:\left[\mathbf{D}(\z,t)f(\z,t)\right],
\end{equation*}
then all the particles belonging to that distribution obey the so-called \emph{Langevin equation},
\begin{equation}
\label{eq:stoc_diff}
d\z\;=\;\left[\dot{\z}+\mathbf{a}(\z,t)\right]dt+\bm{\sigma}(\z,t)\cdot \bm{\Gamma}dt^{1/2},
\end{equation}
where the matrix $\bm{\sigma}$ is defined via a decomposition of the diffusion tensor 
\begin{equation}
\label{eq:basic_decomposition}
2\mathbf{D}=\bm{\sigma}\bm{\sigma}^{T},
\end{equation}
and the upper index $^{T}$ denotes a transpose of a matrix.

The Langevin equation is not a regular differential equation because it contains a \emph{random} variable $\bm{\Gamma}$ zero mean and unit variance. Therefore the Langevin equation is a \emph{stochastic differential equation}. Rewriting the last term as $d\bm{\beta} = \bm{\Gamma}dt^{1/2}$, it stands for an infinitesimal change in the so-called Wiener process $\bm{\beta}$ that now has zero mean and variance $t$, and Langevin equation becomes
\begin{equation}
\label{eq:stoc_diff}
d\z\;=\;\left[\dot{\z}+\mathbf{a}(\z,t)\right]dt+\bm{\sigma}(\z,t)\cdot\bm{\beta}.
\end{equation}
\eero{For an introduction to stochastic processes, see~\citep{oksendal2003stochastic}.} 

Next, we prove that solving the Langevin equation is equal to solving the Fokker-Planck equation. The proof relies on a few basic rules of the so-called \emph{It\^{o} calculus} that extends the methods of ordinary calculus to stochastic processes.

Let us consider an arbitrary function $\phi$ of the stochastic process $\z$ that obeys the Langevin equation and form a differential of it, keeping only the first two terms:
\begin{equation*}
\phi(\z+\Delta\z)-\phi(\z)=\D{}{\z}\phi\cdot \Delta\z+\frac{1}{2}\D{}{\z}\D{}{\z}\phi : \Delta\z \Delta\z+\mathcal{O}(\Delta\z \Delta\z \Delta\z),
\end{equation*}
Without randomness ($\bm{\sigma}=0$), we have $\Delta\z\sim\Delta t$. In that case, dividing by $\Delta t$ and taking the limit $\Delta t\rightarrow 0$, one obtains
\begin{equation*}d\phi=\D{}{\z}\phi\cdot d\z.
\end{equation*}
When the random motion ($\bm{\sigma}\ne 0$) is included, the second order term $\Delta\z \Delta\z$ no longer vanishes because the variance of $\Delta\bm{\beta}$ is proportional to $\Delta t$ and not to $\Delta t^2$. Based on this argument, It\^{o} generalized the rules for stochastic calculus with simple expressions for the mixed differentials 
\begin{equation}
d\z dt=\mathbf{0},\quad d\bm{\beta}dt=\mathbf{0},\quad d\bm{\beta}d\bm{\beta}=\bm{I}dt.
\end{equation}

Using the rules for stochastic differentiation, we find
\begin{align}
\begin{split}
d\phi&=\D{}{\z}\phi\cdot d\z+\frac{1}{2}\D{}{\z}\D{}{\z}\phi : d\z d\z\\
&=\D{}{\z}\phi\cdot\left[(\dot{\z}+\mathbf{a})dt+\bm{\sigma}\cdot d\bm{\beta}\right]+\D{}{\z}\D{}{\z}\phi : \mathbf{D}dt.
\end{split}
\end{align}
Specifying the initial condition as $\z(0)=\mathbf{y}$, the value of the function $\phi(\z)$ at later times can be calculated as
\begin{equation}
\label{eq:phi_change}
\begin{split}
\phi(\z)=&\phi(\mathbf{y})+\int_0^{t}d\phi\\
=&\phi(\mathbf{y})+\int_0^{t}\left(\D{}{\z}\phi\cdot(\dot{\z}+\mathbf{a})+\D{}{\z}\D{}{\z}\phi:\mathbf{D}\right)dt+\int_0^t\D{}{\z}\phi\cdot\bm{\sigma}\cdot d\bm{\beta}.
\end{split}
\end{equation}
Taking the expectation value of Eq.~(\ref{eq:phi_change}) with respect to $\z$ gives the so-called Dynkin's formula, a kind of stochastic generalization of the fundamental theorem of calculus,
\begin{equation}
\mathrm{E}^{\z}\lbrack\phi(\z)\rbrack=\phi(\mathbf{y})+\mathrm{E}^{\z}\lbrack\int_0^{t}A\phi(\z)dt\rbrack,
\end{equation}
where the linear operator $A$, often called the \textit{generator} for the stochastic process $\z$, is defined as
\begin{equation}
A\phi(\z)=\D{}{\z}\phi\cdot(\dot{\z}+\mathbf{a})+\D{}{\z}\D{}{\z}\phi:\mathbf{D}.
\end{equation}
It is worth noting that the contribution from the stochastic term vanishes:
\begin{equation*}
\mathrm{E}^{\z}\lbrack\int_o^t\D{}{\z}\phi\cdot\bm{\sigma}\cdot d\bm{\beta}\rbrack=0.
\end{equation*}

We now recall that the stochastic variable $\z$ obeys a probability density $f(\z,t)$, and the expectation value is also given by
\begin{equation}
\mathrm{E}^{\z}\lbrack\phi(\z)\rbrack=\int\phi(\z)f(\z,t)d\z.
\end{equation}
Then Dynkin's formula can be written as
\begin{equation}
\int\phi(\z)f(\z,t)d\z=\phi(\mathbf{y})+\int\int_0^{t}A\phi(\z)f(\z,t)dtd\z.
\end{equation}
Taking a time derivative of both sides gives
\begin{equation}
\begin{split}
\int\phi(\z)\D{}{t}f(\z,t)d\z=\int A\phi(\z)f(\z,t)d\z\underbrace{=}_{\text{part. int.}}\int\phi(\z)A^{\dagger}f(\z,t)d\z
\end{split}
\end{equation}
where $A^{\dagger}$ is the \textit{adjoint} of the generator $A$
\begin{equation}
A^{\dagger}f(\z)=-\D{}{\z}\cdot\left[(\dot{\z}+\mathbf{a})f\right]+\D{}{\z}\D{}{\z}:\left[\mathbf{D}f\right].
\end{equation}
Rearranging the terms yields
\begin{equation}
\int\phi(\z)\left(\D{}{t}f(\z,t)-A^{\dagger}f(\z,t)\right)d\z=0,
\end{equation}
and, since the function $\phi$ was arbitrary, it holds that 
\begin{equation}
\D{}{t}f(\z,t)=A^{\dagger}f(\z,t)=-\D{}{\z}\cdot\left[(\dot{\z}+\mathbf{a})f\right]+\D{}{\z}\D{}{\z}:\left[\mathbf{D}f\right]
\end{equation}
which is the Fokker-Planck equation.

We have now an alternative means of solving the Fokker-Planck equation: we take an ensemble of test particles, follow their stochastic trajectories integrating the Langevin equation with Monte Carlo mehtods and, in the end, construct the corresponding distribution function as a statistical average of the test particle trajectories. \eero{This sampling method has the additional advantage over more direct methods that complex boundaries of the simulation region are easier to handle. Also many complicated source and sink terms, e.g., particles from nuclear reactions, can be handled in a straight-forward manner.}

In preparation for the guiding-center transformation, let's now have a closer look at the different parts of the Fokker-Planck equation.

\section{Particle phase-space Fokker-Planck equation}
\label{sec:FPcoefficients}
One could simply use the Lorentz force to write $\dot{\z}$. However, here a more formal approach is taken and the particle motion is formulated in terms of a non-canonical Poisson bracket. This is done because, as will be shown, the collisional part of the Fokker-Planck equation can be expressed in terms of non-canonical Poisson brackets. This in turn will be important for the guiding-center transformation of the collision operator.

\subsection{The Hamiltonian motion}
\label{subsec:hamiltonian_motion}
From classical mechanics (for an excellent reference, see, e.g., the book by~\citet{arnold1989}) one recalls that the dynamics of a system is contained in the \emph{Lagrangian one-form}, $\gamma = \mathbf{P}\cdot d\xb - Hd\tau$, where $\mathbf{P}$ is the canonical momentum, $H$ serves as the Hamiltonian constraining the motion of coordinates $(\xb,\mathbf{P})$ along some path, and $\tau$ refers to the path parameter, often measured as real time. The equations of motion for the coordinates are obtained by minimizing the \emph{action integral}, $\int\gamma$. Using \emph{non}-canonical coordinates $z^{\alpha}=(\xb,\vb)$ does not really change the situation. One merely rewrites the canonical momentum as $\mathbf{P}=q\mathbf{A}+m\vb$, or more precisely 
\begin{equation}
\mathbf{P}\cdot\xb=\gamma_{\alpha}dz^{\alpha},
\end{equation} 
and again minimizes the action integral. \eero{Here, summation over repeated indices is assumed.} This leads to equations of motion 
\begin{equation}
\label{eq:eqm-poisson}
\dot{z}^\alpha = \{z^\alpha ,H\}
\end{equation}
where $z^\alpha$ is any of the six phase-space coordinates (eight for time-dependent fields), and on the right-hand-side we have the \emph{Poisson bracket},
\begin{align}
\label{eq:poisson_bracket_divergence_form}
\lbrace f,g\rbrace\equiv\D{f}{z^{\alpha}}\Pi^{\alpha\beta}\D{g}{z^{\beta}},
\end{align}
that is defined in terms of the \emph{Poisson matrix} $\Pi^{\alpha\beta}$ that is the inverse of the \emph{Lagrange matrix}
\begin{align}
\omega_{\alpha\beta}=\frac{1}{2}\left(\D{\gamma_{\beta}}{z^{\alpha}}-\D{\gamma_{\alpha}}{z^{\beta}}\right).
\end{align}

In the absense of time-dependent electric and magnetic fields, so that $\mathbf{E} = -\nabla\Phi$, $\mathbf{B} = \nabla\times\mathbf{A}$, one can use time as the path parameter. The Hamiltonian and Lagrangian are then given by
\begin{align}
H \;=\; & \frac{1}{2} mv^2 + q\Phi,\\
\gamma \;=\; &\left(m\mathbf{v}+q\mathbf{A}\right)\cdot\mathrm{d}\xb-H\mathrm{d}t,
\end{align}
and the Poisson bracket becomes
\begin{align}
\lbrace f,g\rbrace=\frac{1}{m}\left(\nabla f\cdot\D{g}{\mathbf{v}}-\D{f}{\mathbf{v}}\cdot\nabla g\right)+\frac{q\mathbf{B}}{m^2}\cdot\D{f}{\mathbf{v}}\times\D{g}{\mathbf{v}}.
\end{align}
Then it is easy to explicitly verify that Eq.~(\ref{eq:eqm-poisson}) leads to
\begin{align}
\dot{\xb} \;=\; & \vb,\\
\dot{\vb} \;=\; &\frac{q}{m}(\mathbf{E} + \vb\times\mathbf{B}),
\end{align}
which are the familiar equations of motion. The reader should note that allthough the time-dependence of the electromagnetic fields was dropped here, it can be included into the Poisson bracket formalism by extending the six dimensional phase-space $(\xb,\vb)$ to eight dimensions $(\xb,\vb,w,t)$ where $w$ is the total energy of the particle. The Hamiltonian constraint then becomes $H-w=mv^2/2+q\Phi-w$, and the path parameter is $\tau$ instead of the time $t$. This approach also adds one more term to the {\em symplectic} part $\gamma_{\alpha}dz^{\alpha}$ of the Lagrangian. The key is to have a Hamiltonian constraint that is independent of the path parameter which is now $\tau$.

An additional useful relation delivered by this formalism is that the velocity space derivative of any function can be expressed in terms of the non-canonical Poisson bracket:
\begin{equation}
\label{eq:dfdp}
\D{f}{v^i} = \{x^i,mf\}.
\end{equation}
This will turn out to be very practical when carrying out the guiding-center transformation of the collision operator later on.

\subsection{The Collision operator}
In order to get more specific expressions for the Fokker-Planck coefficients $\mathbf{a}$ and $\mathbf{D}$, we have to distinguish between the particle species suffering the collisions, $a$, and the field particles, $b$, that serve as scatterers. \eero{One must, however, remember that $a=b$ is also possible.} The explicit expressions for the Coulomb friction and diffusion coefficients describing the collisions of species $a$ with field species $b$ have been derived, e.g., in Refs.~\citep{PhysRev.107.1:rosenbluth,ichimaru1973basic}:
\begin{align}
\mathbf{a}_{ab}=&\frac{c_{ab}}{m_a^2}\left(1+\frac{m_a}{m_b}\right)\D{h_b}{\mathbf{v}},\\
\mathbf{D}_{ab}=&\frac{1}{2}\frac{c_{ab}}{m_a^2}\D{}{\mathbf{v}}\D{}{\mathbf{v}}g_b,
\end{align}
where $c_{ab}=q_a^2q^2_b\ln\Lambda/(4\pi\epsilon_0^2)$, $m$ is the particle mass, $q$ is the charge, and $f$ the distribution function. The \textit{Rosenbluth potentials} or \textit{Trubnikov potentials}, $h_b$ and $g_b$, are defined as
\begin{align}
h_{b}(\mathbf{v})=\int d\mathbf{v}'f_b(\mathbf{v}')\frac{1}{\left|\mathbf{v}-\mathbf{v'}\right|},\\
g_{b}(\mathbf{v})=\int d\mathbf{v}'f_b(\mathbf{v}')\left|\mathbf{v}-\mathbf{v'}\right|.
\end{align}
These potentials are related to each other:
\begin{align}
\D{}{\mathbf{v}}\cdot\D{}{\mathbf{v}}g_b=2h_b,
\end{align}
which can be used to find a useful relation between the Fokker-Planck coefficients
\begin{align}
\mathbf{a}_{ab}=\left(1+\frac{m_a}{m_b}\right)\D{}{\mathbf{v}}\cdot\mathbf{D}_{ab}.
\end{align}
The two coefficients can now be combined to yield the \textit{Landau form} of the collision operator:
\begin{align}
C_{ab}\left[f_a\right]=&\D{}{\mathbf{v}}\cdot\frac{1}{2}\frac{c_{ab}}{m_a}\int d\mathbf{v}'\left(\frac{\mathbf{I}}{u}-\frac{\mathbf{u}\mathbf{u}}{u^3}\right)\cdot\left(\frac{f_b(\mathbf{v}')}{m_a}\D{f_a}{\mathbf{v}}-\frac{f_a(\mathbf{v})}{m_b}\D{f_b}{\mathbf{v}'}\right),
\end{align}
where $\mathbf{u}=\mathbf{v}-\mathbf{v}'$ is the relative velocity between colliding and field particles. \eero{The Landau form is useful for proving the density, momentum, and energy conservation (see, e.g., the excellent book by~\citet{helander2005}), but here
as a preparation for the guiding-center transformation, we point out an equivalent form}
\begin{align}
C_{ab}\left[f_a\right]=-\D{}{\mathbf{v}}\cdot\left(\mathbf{K}_{ab}f_a-\mathbf{D}_{ab}\cdot\D{f_{a}}{\mathbf{v}}\right),
\end{align}
where the friction coefficient is now defined slightly differently:
\begin{align}
\mathbf{K}_{ab}=\frac{m_a}{m_b}\D{}{\mathbf{v}}\cdot\mathbf{D}_{ab}=\frac{c_{ab}}{m_a^2}\frac{m_a}{m_b}\D{}{\mathbf{v}}h_b.
\end{align}
 Using Eq.~(\ref{eq:dfdp}), it is simple to express the particle collision operator in terms of the non-canonical Poisson brackets:
\begin{align}
\label{eq:C_poisson}
C_{ab}\left[f\right]=-\lbrace x^i,m_aK_{ab}^if_a-m_a^2D_{ab}^{ij}\lbrace x^j,f_a\rbrace\rbrace.
\end{align}
Then, if one knows how to transform the particle Poisson bracket into the guiding-center Poisson bracket, the transformation of the collision operator is practically done, as will be seen in Sec.~\ref{sec:GC}. But first, more explicit expressions for the Fokker-Planck collision coefficients will be given for special cases, important for fusion applications.

\subsection{Collisional coefficients for \emph{minority} ions with isotropic and Maxwellian background plasmas}

From here on, we are focusing on how to simulate minority ions in a stationary background plasma. We thus assume that the colliding particle $a$ is not part of the field particles and, furthermore, that it will not alter the state of the background plasma. With these assumptions, the method can be  used to study, e.g., the impurity ions as well as fast ion distributions in plasmas that have reached a stationary state. From here on, the self-collisions, $C_{aa}$ will thus be neglected.

Often the background plasma, i.e., the field particles, are assumed to have isotropic velocity dependence, $f_b (\z) = f_b(\mathbf{x},v)$. This is a reasonably good assumption when the field particles consist of the thermal plasma in a tokamak, but would not be applicable if one were to evaluate collisions against, say, the fast ion population generated by neutral beams. For isotropic background plasma the coefficients become simpler:
\begin{align}
\label{eq:prtK}
\mathbf{K}_{ab}=&-\nu_{ab}\;\mathbf{v},\\
\label{eq:prtD}
\mathbf{D}_{ab}=&D_{\parallel,ab}\frac{\mathbf{v}\mathbf{v}}{v^2}+D_{\perp,ab}\left(\mathbf{I}-\frac{\mathbf{v}\mathbf{v}}{v^2}\right),
\end{align}
where the scalar coefficients are defined
\begin{align}
\nu_{ab}=&-\frac{c_{ab}}{m_a^2}\frac{m_a}{m_b}\frac{1}{v}h_{b}'(v),\\
D_{\parallel,ab}=&\;\frac{1}{2}\frac{c_{ab}}{m_a^2}g_{b}''(v),\\
D_{\perp,ab}=&\;\frac{1}{2}\frac{c_{ab}}{m_a^2}\frac{1}{v}g_{b}'(v).
\end{align}
Here, the symbols $\parallel$ and $\perp$ refer to the direction of the particle velocity.

When the background plasma is in thermal equilibrium, which is typically assumed of today's tokamak plasmas, the energy distribution of the field particles is Maxwellian,
\begin{align}
f_b=\frac{n_b}{\pi^{3/2}v_b^3}\exp{\left(-v^2/v_b^2\right)}.
\end{align}
Then the diffusion and friction coefficients have analytic expressions:
\begin{align}
D_{\parallel,ab}(v)&=\frac{1}{2}\frac{c_{ab}}{m_a^2}\frac{n_b}{v}G(v/v_b),\\
D_{\perp,ab}(v)&=\frac{1}{2}\frac{c_{ab}}{m_a^2}\frac{n_b}{v}\left(\mathrm{erf}(v/v_b)-\frac{1}{2}G(v/v_b)\right),\\
\nu_{ab}(v)&=\frac{c_{ab}}{m_a^2}\frac{m_a}{m_b}\frac{n_b}{v_b^2}\frac{G(v/v_b)}{v}.
\end{align}
where the Chandrasekhar function $G(s)$ is defined as
\begin{align}
G(s)=\frac{\mathrm{erf}(s)-\frac{2s}{\sqrt{\pi}}\exp(-s^2)}{s^2},
\end{align}
and $\mathrm{erf}(s)$ is the error function.

\subsection{Particle phase-space Langevin equation}
 Now, also decomposing the diffusion matrix according to Eq.~(\ref{eq:basic_decomposition}) becomes rather simple:
\begin{align}
\bm{\sigma}=\sqrt{2D_{\parallel}}\frac{\mathbf{v}\mathbf{v}}{v^2}+\sqrt{2D_{\perp}}\left(\mathbf{I}-\frac{\mathbf{v}\mathbf{v}}{v^2}\right),
\end{align}
where $D_{\parallel}=\sum_bD_{\parallel,ab}$ and $D_{\perp}=\sum_bD_{\perp,ab}$. By direct calculation, one can verify that $\bm{\sigma}\bm{\sigma}^T=2\mathbf{D}$.

The stochastic differential equation describing the particle motion is constructed according to Eq.~(\ref{eq:stoc_diff}). As the Coulomb collisions only affect the particle's velocity coordinates, the equation for the spatial position reduces to the Hamiltonian motion:
\begin{align}
d\xb=\mathbf{v}dt.
\end{align}
For the velocity, the equation includes also the Coulomb drag and diffusion, and is explicitly
\begin{align}
\begin{split}
d\mathbf{v}=&\left[\frac{q}{m}(\mathbf{E}+\mathbf{v}\times\mathbf{B})-\nu_s\mathbf{v}\right]dt+\left[\sqrt{2D_{\parallel}}\frac{\mathbf{v}\mathbf{v}}{v^2}+\sqrt{2D_{\perp}}\left(\mathbf{I}-\frac{\mathbf{v}\mathbf{v}}{v^2}\right)\right]\cdot d\bm{\beta}^{\mathbf{v}},
\end{split}
\end{align}
where $\nu_s=\sum_b(1+m_b/m_a)\nu_{ab}$.
 
\section{Guiding-center formalism with Lie-transform perturbation methods}\label{sec:GC}
Even though we now have practical means for finding the solution to the Fokker-Planck equation, it is still not optimal for solving it numerically. For many applications, resolving the full gyro motion of a charged particle is not only useless, it is also wasteful as far as computing time is concerned. The Larmor frequency is typically two orders of magnitude larger than the bounce frequency, i.e., the frequency at which the particle -- in an axisymmetric situation -- returns to its original position in the poloidal plane. This means that, when integrating the equations of motion corresponding to the Langevin equation, the time step for following the full gyro motion is about one hundredth of the time step that would be satisfactory if one were to follow only the \emph{guiding-center} of the orbit. This, again, translates into one hundred times more time steps in order to reach the physical simulation time. As we shall see in Sec.~\ref{sec:examples}, sometimes it is necessary to follow the full gyro orbit, but in most cases both the magnetic field, and the plasma profiles are practically constant across the Larmor orbit and then resolving the gyro motion does not bring in any useful information. Even in the case of relatively energetic NBI ions, the Larmor radius is very small compared to the gradient length of the magnetic field and plasma profiles in conventional tokamaks (thus excluding spherical tokamaks such as MAST).

Computationally, getting rid of the rapid gyro motion would thus be very welcome. This idea is not new: \citet{alfven1940} proposed averaging the Lorentz force law over the gyro motion explicitly, and his work was carried further before 1970's, most notably by~\citet{gca1961,northrop:adiabaticMotionReview1963}. The basic idea was to expand the magnetic field in a Taylor series around the guiding-center, and then average the Lorentz force law over the gyro-angle. Unfortunately, this simple formalism comes with serious drawbacks, with the most deleterious one being that the equations of motion do not directly conserve energy or posses a Hamiltonian structure. A further complication is related to obtaining the corresponding collision operator: how to average out the Larmor motion from the collision operator? The energy conservation issue does not pose a problem if one integrates over only a handful of orbits, but if one were to construct, say, the slowing-down distribution of fusion alphas, the reliability of the results would be highly questionable. And as for the technical problem with the collision operator, for decades it has simply been ignored and codes around the world have simply used the collision coefficients presented in Sec.~\ref{sec:FP} for the \emph{particle coordinates}.

Therefore, in order to carry out reliable simulations, it is clear that a more formal and consistent approach to get rid of the gyro motion is useful.

\subsection{From gyro-averaging to guiding-center transformation}
The guiding-center is often misunderstood as the average of the particle position over a single gyration around the fieldline, i.e., $\mathbf{X}\approx\int_0^{2\pi}\mathbf{x}\;d\zeta/(2\pi)$, where $\zeta$ is the gyro-angle. However, a rigorous definition is a {\it coordinate transformation}
\begin{align*}
\mathbf{x}=\mathbf{X}+\bm{\rho},
\end{align*}
where $\bm{\rho}$ is the vector from the guiding-center position to the particle position. Its length, $\rho$, is called the Larmor radius. In principle, for any function we have simple transformation rules
\begin{align*}
&f(\mathbf{x})=f(\mathbf{X}+\bm{\rho})=\sum_n\frac{1}{n!}(\bm{\rho}\cdot\nabla)^nf(\mathbf{X})=\exp{(\bm{\rho}\cdot\nabla)}f(\mathbf{X})\equiv F(\mathbf{X}),\\
&F(\mathbf{X})=F(\mathbf{x}-\bm{\rho})=\sum_n\frac{1}{n!}(-\bm{\rho}\cdot\nabla)^nF(\mathbf{x})=\exp{(-\bm{\rho}\cdot\nabla)}F(\mathbf{x})\equiv f(\mathbf{x}).
\end{align*}

Unfortunately, it is not obvious how to define $\bm{\rho}$, i.e., the transformation parameter. The most straightforward approach would be as follows: define the {\it push-forward} operator that pushes particle mappings to guiding-center mappings: $f(\mathbf{x})=\exp{(\bm{\rho}\cdot\nabla)}f(\mathbf{X})=F(\mathbf{X})$, where $\bm{\rho}$ is evaluated at the guiding-center position $\mathbf{X}$. Accordingly, the {\it pull-back} operator pulls guiding-center mappings back to particle mappings: $F(\mathbf{X})=\exp{(-\bm{\rho}\cdot\nabla)}F(\mathbf{x})=f(\mathbf{x})$, where $\bm{\rho}$ is now evaluated at the particle position $\mathbf{x}$. If we now do the intuitive choice and define $\bm{\rho}=-m\mathbf{v}\times\mathbf{B}/(qB^2)$ and evaluate it in the proper place (guiding-center vs particle position), we run into problems: Integrating the particle orbit in {\it nonuniform} $\mathbf{B}$ and evaluating the guiding-center position according to $\mathbf{X}=\mathbf{x}+m\mathbf{v}\times\mathbf{B}/(qB^2)$ gives a point that oscillates at the same frequency as the particle orbit, i.e., the motion of $\mathbf{X}$ would not be {\it independent of the gyrating motion}.

\subsection{About Lie-transformations}
\eero{It was~\citet{littlejohn:guidingCenterHamiltomianJMP1979,littlejohn:hamiltonianPerturbationTheoryJMP1982,littlejohn:variationalPrinciplesJPP1983}, who first managed to construct the guiding-center equations of motion where the rapid gyro motion was eliminated in such a manner that the equations preserved the Hamiltonian structure similarly as the particle phase-space equations (Hamiltonian trajectories never cross, energy invariant exists). His work was subsequently extended on many occasions. We encourage the reader especially to the paper by~\citet{brizard:guidingCenterTransformationPOP1995}, and to the paper by~\citet{RevModPhys.81.693} (and the references therein) where also the history of the guiding-center theory is outlined in detail. In this contribution, we are mostly following these aforementioned works. 

The basic idea behind modern guiding-center theory is to find the coordinate transformation which, first of all, defines the vector $\bm{\rho}$ from the spatial guiding-center position to the particle's spatial position, and, secondly, that in the new coordinates, both the components of the new symplectic part of the fundamental Lagrangian one-form, i.e.,  $\gamma_{\alpha}$, and the Hamiltonian become independent of the gyro-angle. This will then guarantee that the rapid gyro motion is disconnected from the other phase-space coordinates, as it will not appear in the Poisson tensor. A tool that can accomplish all this is the \emph{Lie-transform perturbation theory}. The mathematics involved is rather tedious and \emph{not} straightforward and, therefore, only the procedure will be outlined here. Those interested in explicit intermediate steps in deriving the guiding-center transformation may find them, e.g., in the dissertation of~\citet{EeroPHD}, and those interested in the geometric properties of the Lie-transformation and the associated so-called \emph{Lie-derivative} we refer to the books by~\citet{flanders1989differential} and~\citet{ralph1978foundations}

A Lie-transformation is defined by the \emph{pull-back} and \emph{push-forward} operators
\begin{align}
\mathcal{T} &= \exp(\mathcal{L}_{G}),\\
\mathcal{T}^{-1} &= \exp(-\mathcal{L}_{G}),
\end{align}
which are generalizations of the intuitive operators discussed above. Now, instead of the expression $\bm{\rho}\cdot\nabla$, we have $\mathcal{L}_{G}$, a Lie-derivative, that is generated by a \emph{vector field} $G$ that specifies the transformation. Therefore $G$ is also called the \emph{generating function} of the transformation. The reason why the ordinary transformation is abandoned is that we need a tool that can opearate on \emph{differential forms}: we will transform the Lagrangian one-form that was discussed in Sec.~\ref{subsec:hamiltonian_motion}. A Lie-transformation, however, is not that exotic compared to the ordinary coordinate translation. A Lie-derivative of a function (functions are zero-forms) reduces to $\mathcal{L}_{G}f(\mathbf{x})=G^{\alpha}\partial f/\partial x^{\alpha}$, which is a \emph{directional derivative} with respect to the vector field $G$. A Lie-transformation is thus closely related to the expression $f(\mathbf{x})=\exp{(\bm{\rho}\cdot\nabla)}f(\mathbf{X})$, and a Lie-transformation of a zero-form is in fact a simple coordinate translation along the direction pointed by the vector field $G$. One could then imagine using Lie-transformations to push and pull functions between the particle and guiding-center coordinates, after the transformation itself, the generating function $G$, is chosen. 

It will, however, turn out that a single transformation is not enough to completely eliminate the gyroangle and, in the Appendix~\ref{app:explicit_lie_transformation}, we explain why. As the transformation from particle coordinates to guiding-center coordinates is a \textit{near-identity} transformation (the distance from particle position to guiding-center position is the length of the Larmor radius which is small compared to any other macroscopic length-scale in a plasma), one can adopt a perturbative approach. In the \textit{Lie-transform perturbation theory}, a sequence of transformations, $\mathcal{T}_n=\exp\left(\epsilon^n\mathcal{L}_{G_n}\right)$, with each individual transformation ordered according to the power of $\epsilon$, is thus used, and the guiding-center transformation is defined as an asymptotic series
\begin{equation}
\mathcal{T}_{gc}^{\pm} = \exp\left(\pm\sum_{n=1}^{\infty}\epsilon^n\mathcal{L}_{G_n}\right).
\label{eq:T_gc}
\end{equation}
The dimensionless parameter $\epsilon$, however, is not the familiar ratio of Larmor radius to the typical length scales in the system. It is only used to keep track of different orders while carrying out the transformations. Thus the proper name for $\epsilon$ is the guiding-center \textit{ordering parameter}.

\subsection{Guiding-center transformation}
 \eero{For a charged particle moving in an external magnetic field, the symplectic part of the Lagrangian one-form is 
\begin{equation}
\gamma_\alpha dz^\alpha = \epsilon^{-1}q\mathbf{A}\cdot d\xb + m\vb\cdot d\xb = \epsilon^{-1}\gamma_0 + \gamma_1,
\end{equation}
where we have already prepared the different order of the terms that will affect the motion of the resulting guiding-center. In principle this preparation is simply renormalization of the electric charge but it will result in clear separation of different time scales in the guiding-center motion: order $\epsilon^{-1}$ for the rapid gyro motion, order $\epsilon^{0}$ for the intermediate motion parallel to the magnetic field, and order $\epsilon$ for the slower motion accross the magnetic field lines, i.e., the drift motion. Note, however, that after the transformation is carried up to desired order, the parameter $\epsilon$ is set to unity and not needed anymore. }

Next, applying the perturbative Lie-transform {\em algorithm} starts a sequence of the transformations for both the coordinates and the Lagrangian not really knowing what the vector fields $G_{n}$ generating the transformation truly are. In fact, the generating functions are determined by requiring that the transformed Lagrangian (the symplectic part $\gamma^{\alpha}$ and the Hamiltonian) is independent of the gyro angle $\zeta$.

The transformation we are looking for, $\mathcal{T}_{gc}: z^\alpha \rightarrow Z^\alpha$, changes particle coordinates $z^\alpha  = (\xb,\vb)$ to guiding-center coordinates  $Z^\alpha = (\mathbf{X},v_\parallel,\mu,\zeta)$, where $\mathbf{X}$ is the guiding-center spatial position, $v_\parallel$ is the velocity parallel to the magnetic field, $\mu$ is the so-called {\em magnetic moment}, and $\zeta$ is the gyro angle. The perturbative transformation was defined as an asymptotic series, see Eq.~(\ref{eq:T_gc}), which up to second order is explicitly given by the sequences
\begin{align}
\mathcal{T}_{gc} =& 1 + \epsilon \mathcal{L}_{G_1} +  \epsilon^2(\mathcal{L}_{G_2} + \frac{1}{2} \mathcal{L}_{G_1}^2)+\dots,\\
\mathcal{T}_{gc}^{-1} =&1 - \epsilon \mathcal{L}_{G_1} -  \epsilon^2(\mathcal{L}_{G_2} - \frac{1}{2} \mathcal{L}_{G_1}^2)+\dots,
\end{align}%
and the transformed Lagrangian one-form $\Gamma$ becomes
\begin{equation}
\Gamma = \mathcal{T}_{gc}^{-1}(\epsilon^{-1}\gamma_0 + \gamma_1-H\,dt) + dS = (\epsilon^{-1}\Gamma_0 + \Gamma_1 + \epsilon\Gamma_2 + ...).
\end{equation}
Here $dS$ is a \emph{gauge function} that can be chosen conveniently so that unwanted terms can be cleaned up. It is analogous to the gauges used in electromagnetics and, just like there, it will not affect the physical results. The generating functions $G_n$ can now be determined by requiring that each $\Gamma_n$ is independent of $\zeta$. After a fair amount of algebra and very clever thinking when choosing the generating functions $G_n$, order by order, one obtains
\begin{align}
\Gamma=&\left( \epsilon^{-1}q\mathbf{A}+mv_{\parallel}{\bf\widehat{b}}-\epsilon\frac{m\mu}{q}\bm{R}^{\star} \right)\cdot\mathrm{d}\mathbf{X}+\epsilon\frac{m\mu}{q}\mathrm{d}\zeta-H_{gc}\mathrm{d}t,\nonumber\\
\equiv &\,\epsilon^{-1}q\mathbf{A}^{\star}\cdot\mathrm{d}\mathbf{X}+\epsilon\frac{m\mu}{q}\mathrm{d}\zeta-H_{gc}\mathrm{d}t
\end{align}
where the guiding-center Hamiltonian, $H_{gc}$ is defined as
\begin{equation}
H_{gc} = \frac{1}{2}mv_\parallel^2 + \mu B +\epsilon^{-1}q\Phi\equiv\frac{1}{2}mv_\parallel^2 +\epsilon^{-1}q\Phi^{\star},
\end{equation}
and the so-called modified gyrogauge field is $\bm{R}^{\star}=\bm{R}+(\tau/2){\bf\widehat{b}}$ with $\mathbf{R}$ being the Littlejohn's gyrogauge field that quarantees the gyrogauge invariance of the theory. Here, $\tau={\bf\widehat{b}}\cdot\nabla\times{\bf\widehat{b}}$ is the magnetic field-line twist and ${\bf\widehat{b}}\equiv\mathbf{B}/B$. \eero{In the Hamiltonian, the charge is renormalized with the ordering parameter similarly as in the symplectic part of the Lagrangian one-form, to separate the meaning of different terms: the $\mathbf{E}\times\mathbf{B}$ drift will be of the same order as the parallel motion and the parallel electric field (if it exists, in the absence of collisions, it is free to accelerate the guiding-center to arbitrary energies), will be ordered stronger than the magnetic mirror force that is proportional to gradient of the field and vanishes in a uniform magnetic field.}

Knowing the guiding-center Lagrangian allows construction of the guiding-center version of the Lagrangian matrix, $\omega_{\alpha\beta} $, the inversion of which gives the guiding-center Poisson matrix, $\Pi^{\alpha\beta}$, needed to construct the guiding-center Poisson brackets, $\{F,G\}_{gc}$. \eero{Explicitly, computation of the Poisson bracket gives}
\begin{align}
\label{eq:gc_poisson}
\{F,G\}_{gc}=&\epsilon^{-1}\frac{q}{m}\left(\D{F}{\zeta}\D{G}{\mu}-\D{F}{\mu}\D{G}{\zeta}\right)\nonumber\\
&+\frac{\mathbf{B}^{\star}}{m B_{\parallel}^{\star}}\cdot\left(\nabla F\D{G}{v_{\parallel}}-\D{F}{v_{\parallel}}\nabla G\right)%
-\epsilon\frac{{\bf\widehat{b}}}{qB_{\parallel}^{\star}}\cdot\left(\nabla F\times\nabla G\right),
\end{align}
where $\mathbf{B}^{\star}\equiv\nabla\times\mathbf{A}^{\star}=\mathbf{B}+\epsilon\frac{mv_{\parallel}}{\mu}\nabla\times{\bf\widehat{b}}-\epsilon^2\frac{m\mu}{q^2}\nabla\times\bm{R}^{\star}$ is the so-called \emph{effective} magnetic field, and $B_{\parallel}^{\star}=\mathbf{B}^{\star}\cdot{\bf\widehat{b}}$.
 \eero{The equations of motion can be readily written down:}
\begin{align}
\dot{\mathbf{X}}=&\{\mathbf{X},H_{gc}\}_{gc}=v_{\parallel}\frac{\mathbf{B}^{\star}}{B_{\parallel}^{\star}}+\frac{{\bf\widehat{b}}}{qB_{\parallel}^{\star}}\times q\nabla\Phi^{\star},\\
\dot{v_{\parallel}}=&\{v_{\parallel},H_{gc}\}_{gc}=-\epsilon^{-1}\frac{q}{m}\frac{\mathbf{B}^{\star}}{B_{\parallel}^{\star}}\cdot\nabla\Phi^{\star} ,\\
\dot{\mu}=&\{\mu,H_{gc}\}_{gc}=0,\\
\dot{\zeta}=&\{\zeta,H_{gc}\}_{gc}=\epsilon^{-1}\Omega+\dot{\mathbf{X}}\cdot\bm{R}^{\star}
\end{align}
where $\Omega = qB/m$ is the {\it Larmor frequency} of the particle.

A couple of important observations on the equations of motion can be made: first and foremost they are all independent of $\zeta$, as was desired, and $\mu$ is an explicit constant of motion in the guiding-center formalism. Furthermore, we haven't lost any information. The gyro-angle $\zeta$ can still be followed if so desired. We now have a powerful tool that allows to follow the \emph{drift orbits}, traced by the guiding-center of a test particle, see Fig.~\ref{fig:drift-orbits}.
\begin{figure}
\label{fig:drift-orbits}
\centering
\includegraphics[width=0.4\textwidth]{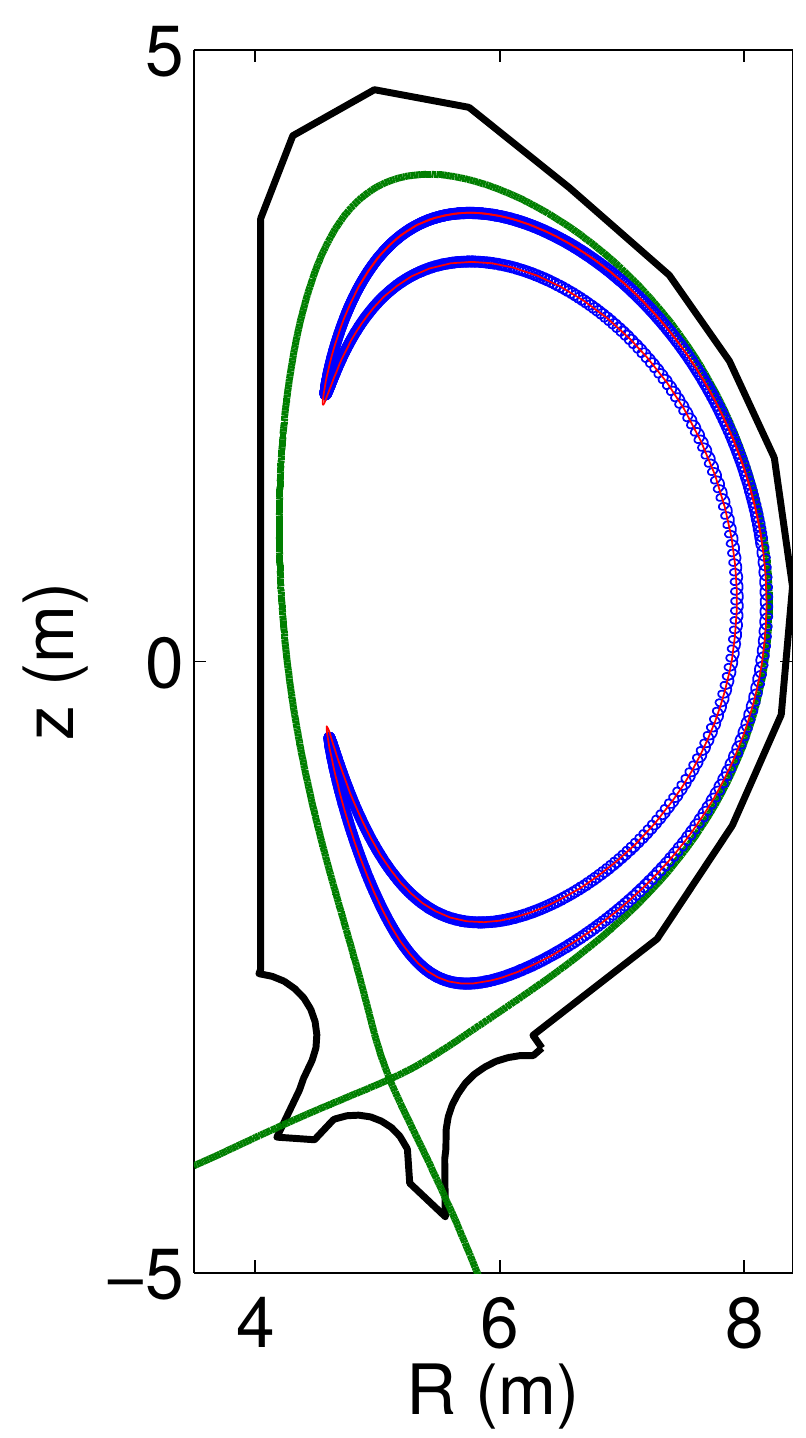}
\includegraphics[width=0.4\textwidth]{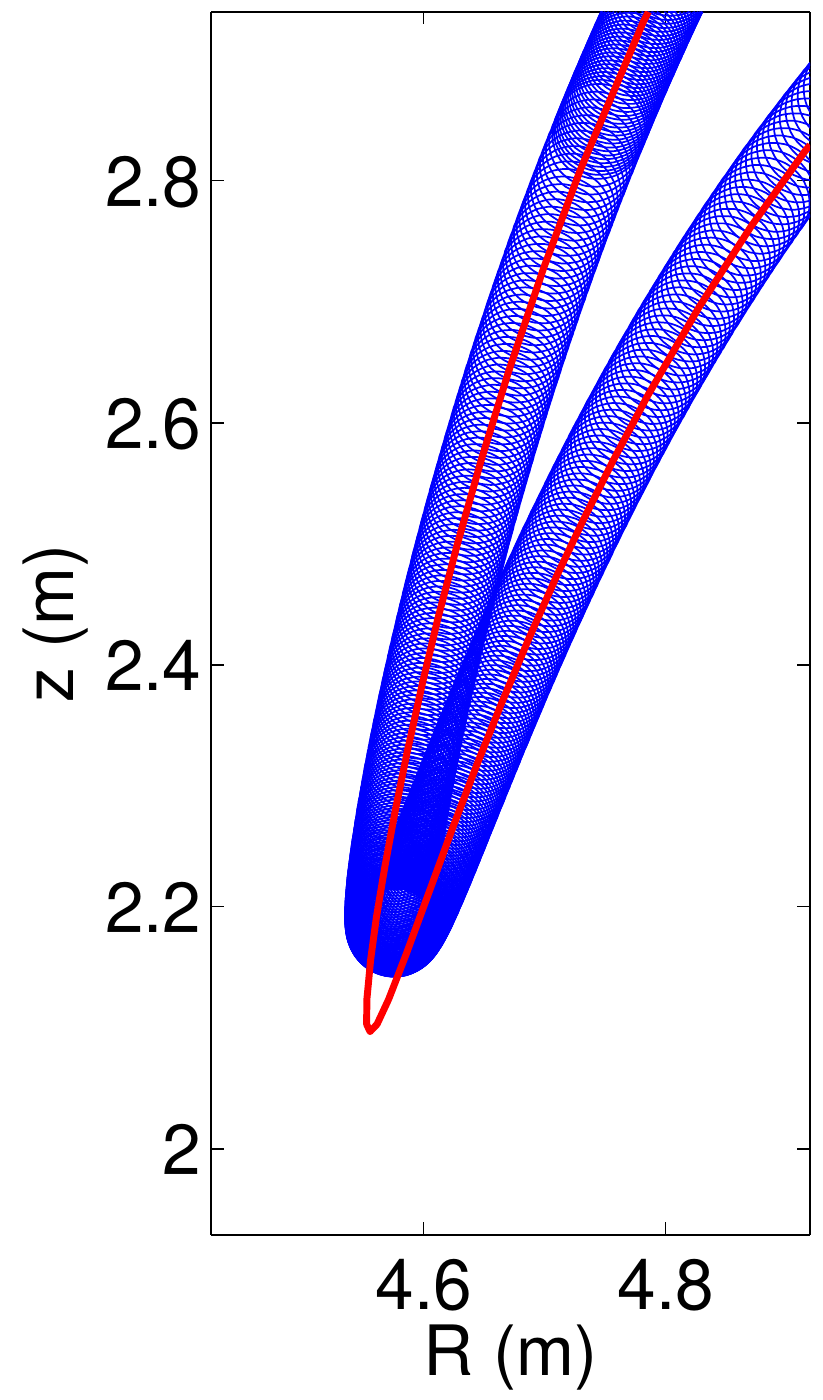}
\caption{\label{fig:orbits}The full gyro orbit (blue) and the corresponding drift orbit (red) traced by the guiding-center of a 3.5~MeV alpha particle in ITER geometry. }
\end{figure}

\subsection{Guiding-center Fokker-Planck and Langevin equations} 
\eero{In this section, for brevity and clarity, we drop the different species labels from the collisional terms. The terms are still assumed to be summations over the different plasma species.

As was discussed in Sec.~\ref{sec:FPcoefficients}, the particle phase-space Fokker-Planck kinetic equation can be written compactly as}
\begin{align}
\D{f}{t}+\{f,H\}=-\{x^i,mK^if-m^2D^{ij}\{x^j,f\}\}.
\end{align}
Transformation into guiding-center phase-space now becomes straight-forward, as we can apply the transformation rule $\{f,g\} \rightarrow \{F,G\}_{gc}$ to obtain
\begin{align}
\label{eq:full_gc_fp}
\D{F}{t}+\{F,H_{gc}\}_{gc}=-\{\mathcal{T}_{gc}^{-1}x^i,m(\mathcal{T}_{gc}^{-1}K^i)F-m^2(\mathcal{T}_{gc}^{-1}D^{ij})\{\mathcal{T}_{gc}^{-1}x^j,F\}_{gc}\}_{gc},
\end{align}
where $F(Z^\alpha)$ is the transformed distribution function in transformed coordinates. %
If we wish to use the Monte Carlo method to solve the guiding-center version of the Fokker-Planck equation, we still need to cast it into a form that allows finding the corresponding Langevin equation. Recalling Sec.~\ref{sec:Langevin}, we have to write the equation in the divergence form:
\begin{align}
\label{eq:gc_kinetic_non_averaged}
\D{F}{t}=-\frac{1}{\mathcal{J}}\D{}{Z^{\alpha}}\left[\mathcal{J}\left(\left(\dot{Z}^{\alpha} + \mathcal{K}^{\alpha}\right)F-\mathcal{D}^{\alpha\beta}\D{F}{Z^{\beta}}\right)\right],
\end{align}
where $\mathcal{J}$ is the phase-space Jacobian of the new coordinate system, and the guiding-center friction and diffusion coefficients, $\mathcal{K}^{\alpha}$ and $\mathcal{D}^{\alpha\beta}$, are given by
\begin{align}
\mathcal{K}^{\alpha}=&m(\mathcal{T}_{gc}^{-1}\mathbf{K})\cdot\bm{\Delta}^{\alpha},\\
\mathcal{D}^{\alpha\beta}=&(\bm{\Delta}^{\alpha})^{\dagger}\cdot m^2(\mathcal{T}_{gc}^{-1}\mathbf{D})\cdot\bm{\Delta}^{\beta},
\end{align}
where $\bm{\Delta}^{\alpha}=\{\mathcal{T}_{gc}^{-1}\xb,Z^{\alpha}\}$ are so-called projection vectors, defined by the guiding-center Poisson tensor~\citep{brizard:fokkerPlanck}.

Before rushing into writing out the guiding-center Langevin equation, it should be remembered that the guiding-center distribution function $F(Z^\alpha)$ as well as the friction and diffusion coefficients still contain the $\zeta$ dependence, so far only the equations of motion, $\dot{Z}^{\alpha}$ are independent of the gyro-angle. Therefore, the Fokker-Planck equation first has to be averaged over $\zeta$. \eero{As the collisional time scale in typical fusion plasmas is much longer than the gyro motion time-scale,~\citet{brizard:fokkerPlanck} managed to construct a closure scheme that gives the guiding-center collision operator in the reduced 5D guiding-center phase-space.} The resulting reduced guiding-center Fokker-Planck equation can be expressed in a form that eventually allows us to identify the necessary coefficients for the Langevin equation~\citep{hirvijoki:092505:2013:pop}
\begin{align}
\label{eq:gcKolmogorovForward}
\frac{\partial \mathcal{F}}{\partial t}=&-\frac{1}{\mathcal{J}}\frac{\partial}{\partial Z^{\alpha}}\left(\mathcal{J}(\dot{Z}^{\alpha}+\langle\mathcal{A}\rangle^{\alpha}) \mathcal{F}\right)+\frac{1}{\mathcal{J}}\frac{\partial^2}{\partial Z^{\alpha}\partial Z^{\beta}}\left(\mathcal{J}\langle\mathcal{D}\rangle^{\alpha\beta} \mathcal{F}\right).%
\end{align}
Here $\mathcal{F}=\left\langle F\right\rangle$ now stands for the gyro-averaged guiding-center distribution function, and the gyro-averaged drift coefficient is defined as
\begin{align}
\label{eq:gcdrift}
\langle\mathcal{A}\rangle^{\alpha}=\langle\mathcal{K}\rangle^{\alpha}+\frac{1}{\mathcal{J}}\frac{\partial}{\partial \mathcal{Z}^{\beta}}(\mathcal{J}\langle\mathcal{D}\rangle^{\alpha\beta}).
\end{align}
Now we are in the position to use the connection to stochastic processes and to obtain the guiding-center Langevin equation that is rid of the fast gyrating motion:
\begin{align}
\label{eq:gcmc}
dZ^{\alpha}=\left(\dot{Z}^{\alpha} + \langle\mathcal{A}\rangle^{\alpha}\right)dt+\Sigma^{\alpha\beta}d\mathcal{W}^{\beta},
\end{align}
where $d\mathcal{W}^\alpha$ is again a Wiener process with zero mean and variance t, and
 $\Sigma^{\alpha\beta}$ can be computed from
\begin{equation}
\mathcal{D}^{\alpha\beta}=\frac{1}{2}\Sigma^{\alpha\gamma}\Sigma^{\beta\gamma}.
\end{equation}

 Unlike in the particle phase-space, decomposing of the matrix $\mathcal{D}^{\alpha\beta}$ is highly non-trivial. For those interested, more details can be found in Refs.~\citep{hirvijoki:092505:2013:pop,EeroPHD}. \eero{As an example, we give the guiding-center Fokker-Planck coefficients in a non-uniform magnetic field, but in case where the plasma density and temperature are constant or vary little across the magnetic field lines:} the gyro-averaged friction coefficients are
\begin{align}
\langle\mathcal{K}^{\mathbf{X}}\rangle&=\epsilon\nu\frac{\mathbf{\widehat{b}}}{\Omega_{\parallel}^{\star}}\times\dot{\mathbf{X}}+\mathcal{O}(\epsilon^3),\\
\label{eq:Kv}
\langle\mathcal{K}^{v_{\parallel}}\rangle&=-\nu v_{\parallel}-\epsilon\lambda\frac{\mu B}{mv_{\parallel}}\nu+\mathcal{O}(\epsilon^2),\\
\label{eq:Ku}
\langle\mathcal{K}^{\mu}\rangle&=-(2-\epsilon\lambda)\nu\mu+\mathcal{O}(\epsilon^2),
\end{align}
where $\lambda=v_{\parallel}{\bf\widehat{b}}\cdot\nabla\times{\bf\widehat{b}}/\Omega$, and the corresponding diffusion coefficients are
\begin{align}
\label{eq:Dxx}
\langle\mathcal{D}^{\mathbf{X}\mathbf{X}}\rangle=&\epsilon^2\left[(D_{\parallel}-D_{\perp})\frac{\mu B}{2\mathcal{E}}+D_{\perp}\right]\frac{\mathbf{I}-{\bf\widehat{b}}{\bf\widehat{b}}}{(m\Omega_{\parallel}^{\star})^2}+\mathcal{O}(\epsilon^3),\\
\label{eq:Dvv}
\langle\mathcal{D}^{v_{\parallel}v_{\parallel}}\rangle=&\frac{D_{\parallel}}{m^2}+(1-\epsilon\lambda)\frac{D_{\perp}-D_{\parallel}}{m^2}\frac{\mu B}{\mathcal{E}}+\mathcal{O}(\epsilon^2),\\
\label{eq:Duu}
\langle\mathcal{D}^{\mu\mu}\rangle=&(1-\epsilon\lambda)\frac{2\mu}{mB}\left[(D_{\parallel}-D_{\perp})\frac{\mu B}{\mathcal{E}}+D_{\perp}\right]+\mathcal{O}(\epsilon^2),\\
\label{eq:Dxv}
\langle\mathcal{D}^{\mathbf{X}v_{\parallel}}\rangle=&\epsilon^2\frac{v_{\parallel}}{(m\Omega_{\parallel}^{\star})^2}(D_{\parallel}-D_{\perp})\frac{\mu B}{2\mathcal{E}}\nabla_{\perp}\ln{B}\nonumber\\&+\epsilon^2\frac{v_{\parallel}}{(m\Omega_{\parallel}^{\star})^2}\left[D_{\parallel}+\frac{\mu B}{2\mathcal{E}}(D_{\perp}-D_{\parallel})\right]\mathbf{\widehat{b}}\cdot\nabla\mathbf{\widehat{b}}+\mathcal{O}(\epsilon^3),\\
\label{eq:Dxu}
\langle\mathcal{D}^{\mathbf{X}\mu}\rangle=&-\epsilon\frac{\mu}{2m\mathcal{E}}(D_{\parallel}-D_{\perp})\frac{\mathbf{\widehat{b}}}{\Omega_{\parallel}^{\star}}\times\dot{\mathbf{X}}+\mathcal{O}(\epsilon^3),\\
\label{eq:Dvu}
\langle\mathcal{D}^{\mu v_{\parallel}}\rangle=&(1-\epsilon\lambda)\frac{\mu v_{\parallel}}{m\mathcal{E}}(D_{\parallel}-D_{\perp})+\epsilon\lambda\frac{\mu}{v_{\parallel}m^2}D_{\parallel}+\mathcal{O}(\epsilon^2),
\end{align}
\eero{If the gradients in the plasma profiles would be accounted for, they would essentially introduce Finite-Larmor-Radius effects into the scalar coefficients $\nu,D_{\parallel},D_{\perp}$. An interested reader may consult Ref.~\citep{decker:orbitAveragedFokkerPlanck} for more details.}

It is clear that the increased efficiency of the computational effort comes at the price of mathematical complexity. The most important feature to notice from the complicated expressions for the diffusion coefficients is that, unlike in the particle picture, in the guiding-center picture the collision operators no longer act only in the velocity space. The collisions in the guiding-center picture change also the position of the \emph{guiding-center itself}, as indicated by the spatial indices in the friction and diffusion coefficients above. This is easily understood, since a change in particle velocity changes its Larmor radius and, thus, its guiding-center position. This is also illustrated in Fig.~\ref{fig:GCcollisions} that shows how applying the same collision in particle frame and guiding-center frame-of-reference produces the spatial step only in the latter.

\begin{figure}
\centering
\includegraphics[width=0.5\textwidth]{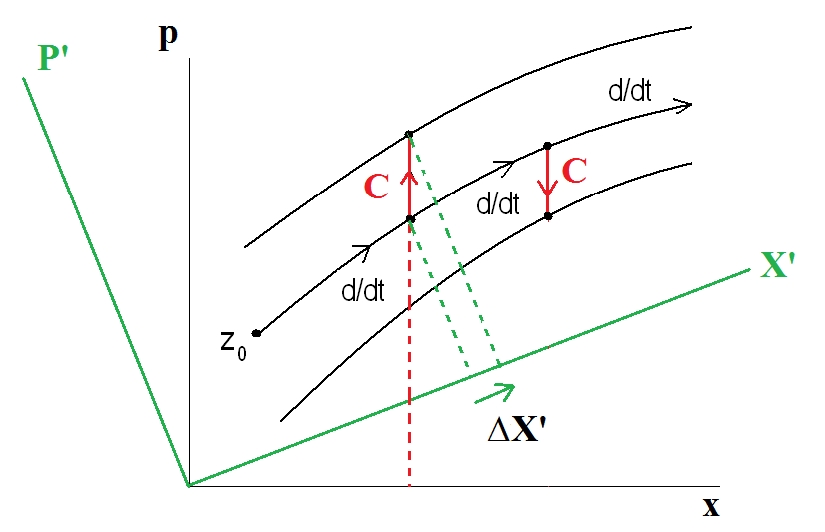}
\caption{\label{fig:GCcollisions} The effect of collisions in particle space and in the transformed guiding-center space, indicated by the primed axes. It is clear that while in the particle space the particle is kicked from one Hamiltonian trajectory to another only in the velocity (momentum) space, in guiding-center picture the collisional kick also introduces a spatial step $\Delta${\bf X}. \emph{Courtesy of Alain Brizard}}
\end{figure}

We now have a powerful tool that allows us not only to follow the drift orbits of charged particles in fusion plasmas but also reproduces \emph{neoclassical transport} of minority populations due to the combined effect of Coulomb collisions and toroidal geometry. This is illustrated in Fig.~\ref{fig:NCorbits}, where the guiding-center orbit changes in time due to collisions: a banana orbit is \emph{transported} into a passing orbit, and vice versa. In the next Section, we discuss when and how supercomputers should be used to answer pressing questions in fusion research that can be tackled with this method, and after that we devote a section to a few examples where the method has been found indispensable.
\begin{figure}
\includegraphics[width=.45\linewidth]{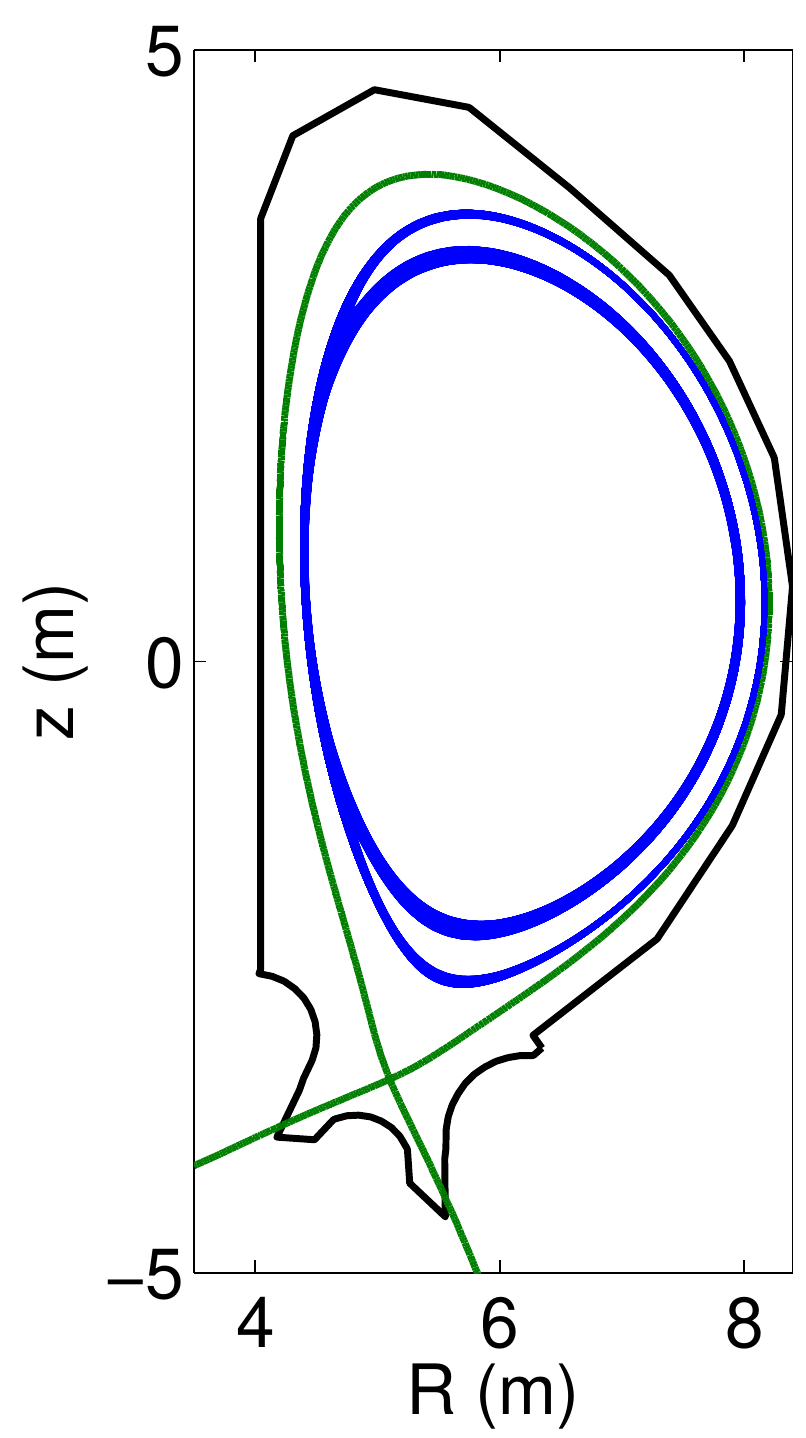}
\includegraphics[width=.45\linewidth]{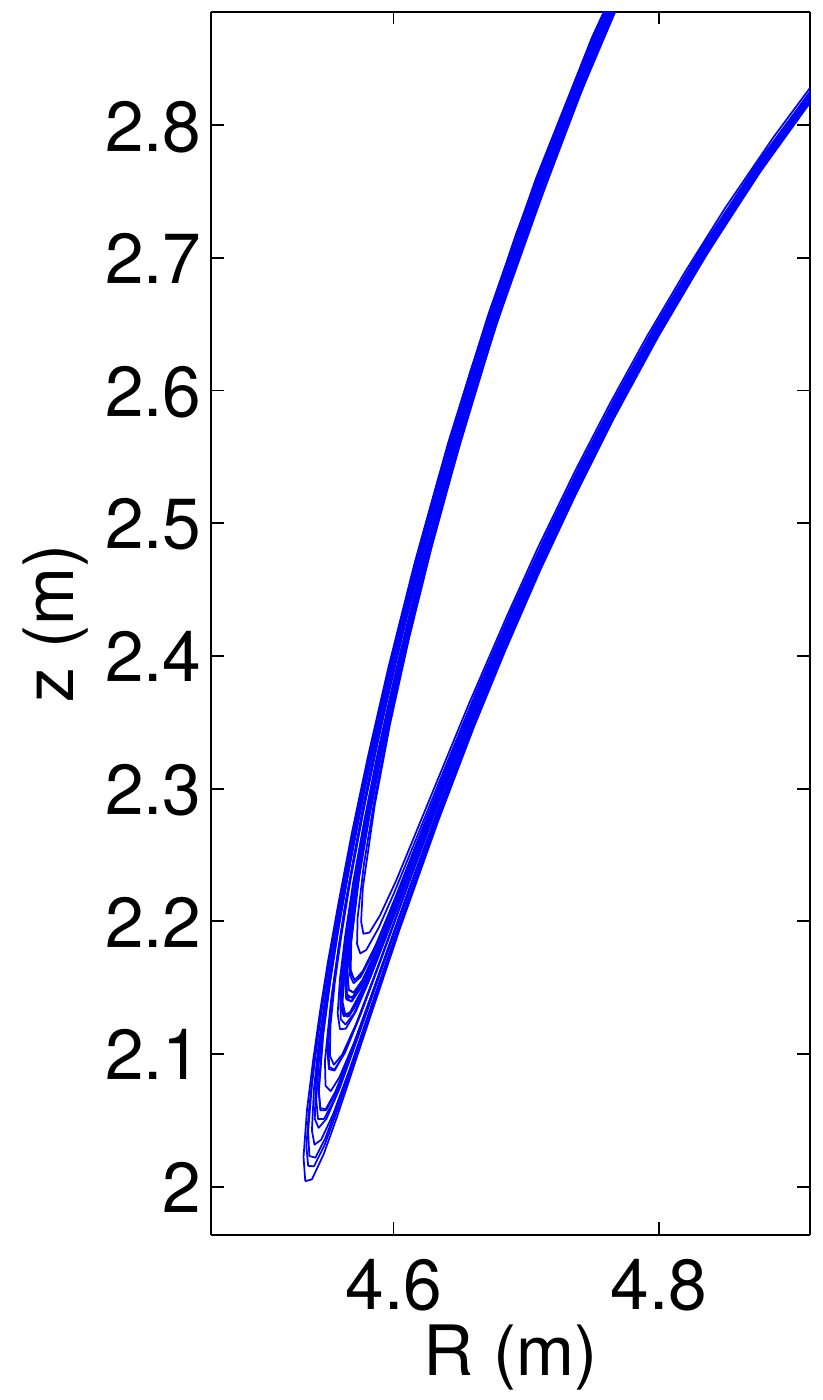}
\caption{\label{fig:NCorbits} (a) A banana orbit traced by the guiding-center method for 100ms in ITER geometry. Unlike in Fig.~\ref{fig:orbits}, here the collisions are included and the orbit no longer closes on itself in the poloidal plane, leading to neoclassical transport, seen as displacements of the banana tips on subsequent orbits (b). Eventually, the originally banana orbit becomes a passing orbit when the upper and lower tips connect.}
\end{figure}

\section{High Performance Computing}\label{sec:HPC}
In this section, we wish to first justify the need for supercomputers, then explain the main differences between using a collection of regular computers or one supercomputer, and finish the section by sketching how codes like ASCOT benefit from using supercomputers.

\subsection{Who needs supercomputers?}
In high performance plasmas, the \emph{magnetic field} dominates the test particle behaviour. How we get the values of the magnetic field and its derivative, needed for the equations of motion, has been evolving over the years. Before the 1990's, it was customary to assume that the plasma had a large aspect ratio, $R/a \gg 1$ and was poloidally circularly symmetric with a simple material limiter. With these assumptions, it was possible to write down an analytical expression for ${\mathbf B}$. This method of simply calculating the field value and its derivative is very fast and very accurate and, furthermore, using \emph{field-aligned} coordinates like Boozer coordinates is straightforward. Unfortunately, introduction of plasma shaping and, in particular, the \emph{separatrix} and the \emph{scrape-off-layer} (SOL) due to divertor operation of the plasmas made this approach practically useless. To take such a non-trivial geometry into account in simulation forced adopting tabulated backgrounds, where the field components (and possibly derivatives as well) are given on a discrete 2-dimensional grid in $(R,z)$ space. For instance, ASCOT got 2D magnetic backgrounds in $\sim$1997, allowing simulations in the SOL and divertor regions. Since the particle motion is not restricted to the grid points, \emph{2D interpolation} is needed. This slows down the simulations significantly. Furthermore, the choice of the interpolation routine is not trivial, but one has to pay a lot of attention not only to its speed but also to its accuracy and smoothness.

The remaining symmetry is given by the assumption that tokamaks are axisymmetric devices. This is a reasonable assumption for, e.g., JET, that has 32 toroidal field coils. Unfortunately, it breaks badly down for devices like ITER, where the sparseness and limited extent of the 16 toroidal field coils makes the toroidal magnetic ripple unacceptably large for confining energetic ions, see Fig.~\ref{fig:ripples}. Therefore ITER is equipped with \emph{ferritic inserts} that effectively reduce the ripple as indicated in Fig.~\ref{fig:ripples}. However, the situation is worsened by the presence of other ferritic components, such as the Test Blanket Modules (TBM) used to test tritium breeding in ITER, and external coils to mitigate the ELMs. The total destruction of axisymmetry in the presence of these perturbations is obvious from Fig.~\ref{fig:ripples}. Therefore, for devices like ITER, let alone stellarators, we need the full 3-dimensional magnetic field with the need for 3D interpolation. For instance, support for experimental 3D backgrounds was implemented in ASCOT around 2007.

\begin{figure}
\centering
\includegraphics[width=0.5\textwidth]{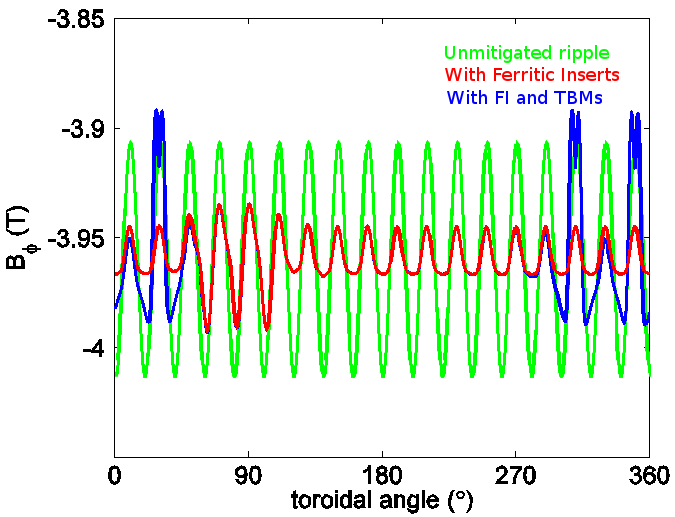}%
\caption{\label{fig:ripples}The strength of the toroidal magnetic field in ITER as a function of toroidal angle in a single $(R,z)$ point. The different colors correspond to different magnetic fields: The green one is the field only due to the toroidal field coils. The red line shows how the ferritic inserts reduce the toroidal ripple amplitude. The NBI beams limit the ferritic insert size near 90$^\circ$, which explains the larger ripple there. The blue line shows the complete field, including the perturbation due to the ferromagnetic TBMs.}
\end{figure}

Simulating energetic ions in full 3-dimensional field brings us to the domain of supercomputing. For instance, a 3.5~MeV alpha-particle in ITER slows down in $\sim$1\,s. Simulating it in realistic 3D magnetic field, including collisions and a realistic description of the machine first wall etc, takes $\sim$1\,min when using the guiding centre approach, or  $\sim$1\,h when following the full gyro orbits. Depending on the quantity of interest, one needs from thousands to millions of test particle alphas and, therefore, super computers become very attractive. What follows applies to codes that are run with some thousands of CPUs. Codes requiring more CPUs would use more advanced tools.

\subsection{A surgical cut to the intestines of a supercomputer}
The computational speed of individual computing devices today is limited by power and cooling requirements. Simulating large systems consequently requires parallel computing, in which the overall problem is divided into subtasks that can be computed concurrently on numerous devices working in parallel.

The computational effort in a modern supercomputer is parallelized on multiple levels (Fig.~\ref{fig:supercomputer}). A typical supercomputer is a cluster consisting of thousands of individual computers, called \emph{nodes}, connected through a fast network. Each node is an independent computer typically running a Linux operating system. It contains 2-4 central processing units (CPU), each of which consists of 4-16 parallel computing units or \emph{cores}. Furthermore, each core is capable of processing machine instructions from multiple parallel \emph{threads}, meaning copies of the code running concurrently. Finally, so-called SIMD (Single Instruction, Multiple Data) instructions can perform the same operation on multiple data elements simultaneously, greatly increasing the throughput of the core.
\begin{figure}\label{fig:supercomputer}
\centering
\includegraphics[width=0.5\textwidth]{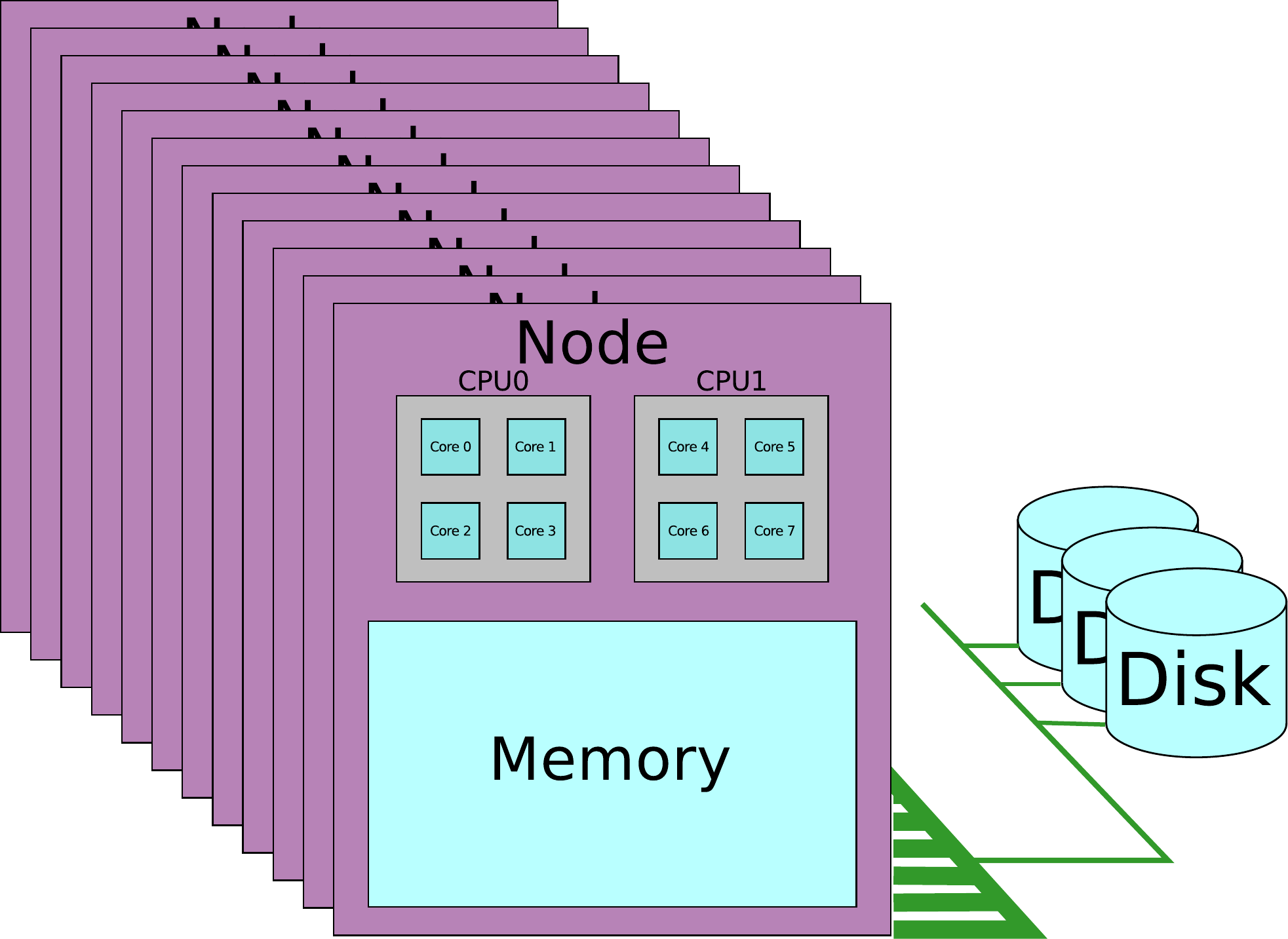}%
\caption{Schematic picture of typical supercomputer components. Inside a node, the CPUs have very fast access to the memory of that node, but access to the different memories outside the node slow down quickly as indicated by the thickness of the green routes: the nodes can still communicate with each other at reasonable speed, but one should avoid accessing data on the disks since that connection is very slow.}
\end{figure}

In addition to the CPU's, modern supercomputer nodes can include so-called \emph{accelerators} or \emph{coprocessors}. They contain a large number of parallel cores, which are individually simpler and slower than the cores in the CPU, but which can efficiently perform parallel tasks. A typical mode of operation is to offload well parallelizable computations to the accelerator device and process the results in the CPU.

A user on a supercomputer normally does not directly work on the computing nodes. Instead, separate login nodes are provided for user access. The computing tasks are submitted to a scheduling system, which handles the actual execution of the code on the computing nodes. File access is also typically provided through separate storage nodes, which the computing nodes can access over the network.

The limiting factor in modern supercomputers is typically communication between the numerous parallel processes, which requires low latency and high bandwidth data transfer. The processes within a single node often can all access the same memory space in a shared memory environment, which enables very fast communication. However, processes running on separate nodes, a distributed memory environment, must use the network layer for communication, which is typically orders of magnitude slower than communicating through shared memory.  This is emphasized in so-called computing grids, in which the nodes may be connected through an ordinary local network or the Internet, with limited bandwidth and high latency. The following section describes how these issues influence Monte Carlo orbit-following codes.

\subsection{Applying supercomputers to Monte Carlo orbit-following codes}

Solving the distribution through Monte Carlo orbit-following involves simulating the trajectories of test particles and gathering their distribution over time. An initial state for the test particles is generated, for example representing the distribution of alpha particles produced in fusion reactions. The stochastic differential equations describing the motion of the particles are integrated numerically, which requires numerical representations for the background fields and plasma and the geometry of the simulated region. Desired parameters of the particles are computed at each step and gathered in a multidimensional histogram, from which the distributions are computed at the end of the simulation.

In minority particle orbit-following, the behaviour of each test particle is independent of the others. Consequently, the processes simulating the particles in principle need not communicate with each other during the simulation run. Only at the end of the simulation the distributions of the individual particles must be combined together to form the final distribution. This is an example of an \emph{embarrassingly parallel} problem which can easily be computed at a massively parallel scale, and which is especially suitable for grid computing. At the most basic level, this would involve executing copies of the code, each producing a distribution for an individual particle, on each processor in the computing system and combining the results.

In practice, several factors may still limit the performance of the code. The magnetic field structure, multi-dimensional distributions and other simulation parameters require a considerable amount of memory, which enforces a limit on the number of independent processes that can be executed in parallel on a single node due to the limited memory capacity. Single copies of the data structures suffice when simulating multiple particles in a shared memory environment, but this in turn brings out issues with concurrent memory accesses. In particular, updating the distributions requires synchronization of the parallel threads, incurring a performance penalty, to avoid data loss from race conditions, in which a thread attempts to update the distribution simultaneously with another thread.

Another factor that can affect the performance is file access. As the number of storage nodes in a supercomputer is limited, reading and writing the data separately in each process can have a significant impact on the execution time. To avoid this, file access can be centralized to a separate process, which distributes the data to and gathers the results from the computing processes.

A number of tools exist for efficiently running a code in a parallel environment. Message Passing Interface (MPI) is a commonly used interface for executing multiple processes over multiple nodes and communicating data between them. File access can be implemented through libraries such as HDF5 or NetCDF, which provide parallelized reading and writing and compressed data transfer, reducing the bandwidth requirements. Computing grids can be created with tools such as HTCondor, which can be used to run computations on workstations when they are otherwise unoccupied. On each individual node, interfaces such as OpenMP or OpenACC can be used to execute the code concurrently in a shared memory environment. These provide a simple method for specifying the concurrent segments of the code, which the compiler then automatically parallelizes without the need for explicitly implementing the code for each parallel thread separately. The next section describes through examples how these tools and techniques are applied in practice in the case of ASCOT orbit-following code.

\section{Applying the formalism}\label{sec:examples}
In this section we give a few representative examples of cases where the formalism outlined in this paper been applied. All cases have 3-dimensional magnetic fields and realistic 3-dimensional wall configurations, which renders the problems essentially impossible for any other means than Monte Carlo methods. All simulations have been carried out using ASCOT, a suite of codes developed at the Aalto University. We start by introducing the most important features of ASCOT, together with some key numbers related to its performance. This is followed by three examples, representing different features that make Monte Carlo approach attractive.

\subsection{ASCOT -- race track for tokamak particles}
ASCOT is actually a suite of codes. The highly parallel, computationally intensive parts are written in FORTRAN 90, while many preprocessing, postprocessing, analysis and visualisation tools are written in MATLAB. The key components include particle generators, the main orbit following routine including operators representing various interactions and wall-collision routines. Post-processing tools include models for synthetic diagnostics. For particle generation, ASCOT features ab initio particle loading for fusion alphas from thermo-nuclear and beam-target reactions, as well as ions generated by neutral beam injection. These ions can be followed either by integrating their full gyro motion (necessary, e.g., for beam ions in MAST), or using the guiding-center approach. ASCOT also provides a hybrid method, where the guiding-centers of the ions are followed until the ion comes into the vicinity of a material surface. At this point, the ion is shifted to its Larmor orbit and the full gyro motion is traced parallel to the guiding-center motion. If the ion does not collide with the wall but starts receding from it, the full orbit following is simply dropped. As for interactions, in addition to the collision operators introduced above, ASCOT features a unique numerical model for MHD modes of NTM (almost stationary) and TAE (rapidly rotating) type~\citep{mhdpapru}. This model is applicable in arbitray coordinate system and thus allows extending the simulations across the separatrix. Furthermore, ASCOT can also take into account the effect of SOL flows. As for diagnostics, a 4-dimensional, $(R,z,E,\xi)$ or $(R,z,v_\parallel ,v_\perp)$, distribution is collected, along with several one dimensional profiles and two dimensional distributions of many quantities of interest, such as power deposition, electrical current or torque exerted by the fast ions. For those interested, more details of ASCOT can be found in Ref.~\citep{ascot4ref}.

\subsection{How wall can make a world of difference}
The 3.5~MeV alphas, born in large quantities in the fusion reactions, could jeopardize the integrity of the ITER first wall and lead to cooling water accidents that are expensive and time-consuming to repair. Therefore their confinement is of key importance, but also the structure of the wall plays a role. Over the years, ITER design has undergone several revisions, and the design of the first wall is no exception. Initially, the first wall was foreseen to exhibit two limiters, placed toroidally opposite to each other as shown in the upper row of Fig.~\ref{fig:ITERwall}. Figure~\ref{fig:ITERwall}(a) shows the power load on the wall due to fusion alphas, calculated under the assumption of axisymmetric field, while Fig.~\ref{fig:ITERwall}(b) shows the same for unmitigated toroidal ripple. The ripple increases the peak power loads to alarming levels. When the effect of the FIs is included in Fig.~\ref{fig:ITERwall}(c), the calculated power loads are practically indistinguishable from the axisymmetric case. Figure~\ref{fig:ITERwall}(d) shows the fusion alpha power load on the ITER first wall with the most recent limiter construction, so-called \emph{continuous} limiter, where limiter plates are placed at each of the 16 toroidal sectors.

The ripple is known to be a risk for the ITER operation and, hence, ITER is equipped with \emph{ferritic inserts} (FI)that reduce the toroidal ripple. When the effect of the FIs is included, the calculated power loads are practically indistinguishable from the axisymmetric case. Figure~\ref{fig:ITERwall}(c) shows the fusion alpha power load on the ITER first wall with the most recent limiter construction, so-called \emph{continuous} limiter, where limiter plates are placed at each of the 16 toroidal sectors. This configuration clearly spreads out the power load to very benign levels.

Comparing the three pictures in Fig.~\ref{fig:ITERwall} makes it clear that with the wall construction one can greatly influence the power loads due to energetic particles.

\begin{figure}
\includegraphics[width=.48\linewidth]{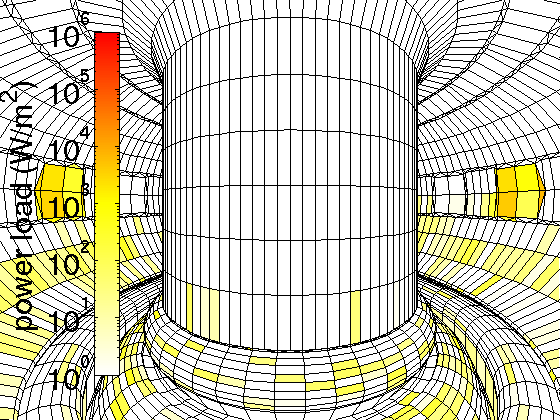}
\hfill
\includegraphics[width=.48\linewidth]{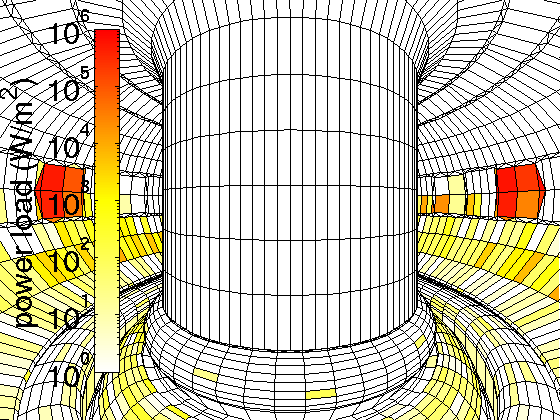}

\vspace{4mm}

\includegraphics[width=.48\linewidth]{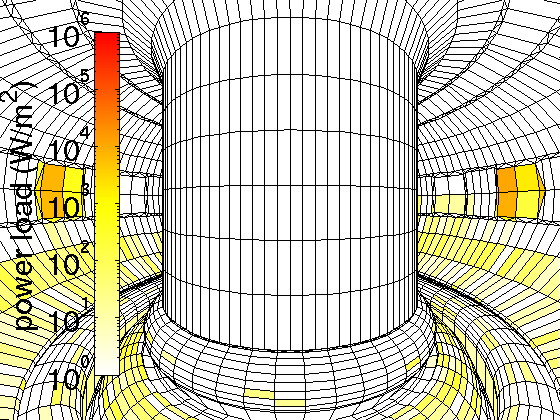}
\hfill
\includegraphics[width=.48\linewidth]{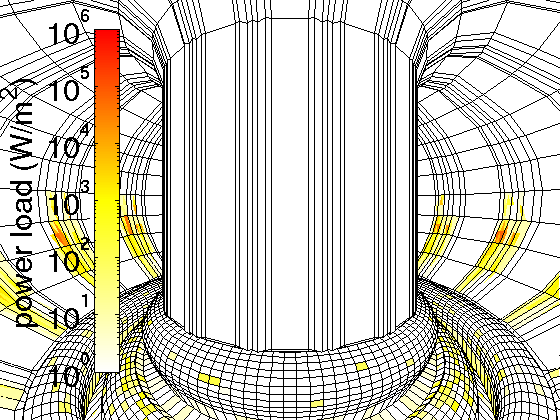}
\caption{\label{fig:ITERwall} The simualted power load due to lost alpha particles on the first wall of ITER in the 9MA advanced scenario. (a) The power load assuming an axisymmetric magnetic field, (b) the power load assuming unmitigated ripple, (c) the power load assuming ripple mitigated by ferritic inserts and (d) the power load assuming a continuous limiter structure (with ripple mitigation). Note the color scale is logarithmic.}
\end{figure}

\subsection{TBM-mockup experiments at DIII-D}\label{online}
As explained in Sec.~\ref{sec:HPC}, the axisymmetry of the ITER magnetic field will be compromised by the TBMs. Since lack of axisymmetry is directly translated into reduced confinement, alarm was sounded on possible wall damage due to localized losses of energetic alphas and/or NBI ions near the TBMs. This led to not only significant simulation efforts, but even experimental studies where the effect similar to that due to TBMs was produced by introducing mock-up coils behind the first wall tiles in the DIII-D tokamak~\citep{Kramer13_fast_ion_heat_loads_TBM}.

ASCOT was used to simulate the confinement of NBI ions when the mock-up coils were activated. The power load on the first wall is illustrated in Fig.~\ref{fig:d3d-wall}. The highest power is received by not only the two protruding limiters, but on four tiles between them. These are exactly the tiles behind which the mock-up coils were located, and they were also found to heat up during the experiment. Experimentally, the loss of beam ions was also reflected in the neutron signal: even in deuterium plasma, a finite number of DD reactions take place between the fast beam ions and the thermal plasma. As a result, tritium appears in the plasma, and subsequent DT reactions produce the characteristic 13.5~MeV neutrons in large enough numbers to be experimentally measured. When the mock-up coils were turned on, the neutron signal dropped about 30\%~\citep{kramer:nf:51:103029}. ASCOT was used to simulate the tritium born in the plasma, in this case using the option of following the full gyro motion of the particles. The need for this is indicated in Fig.~\ref{fig.d3d-T}(a) that shows the orbit of the 1~MeV triton. Two simulations were carried out: one with and one without the effect of the mock-up coils. The number of confined tritons, normalized to the case without the coils, from the two cases is shown in Fig.~\ref{fig.d3d-T}(b), showing quite good agreement with the experimental observation.

\begin{figure}
\centering
\includegraphics[width=.45\linewidth]{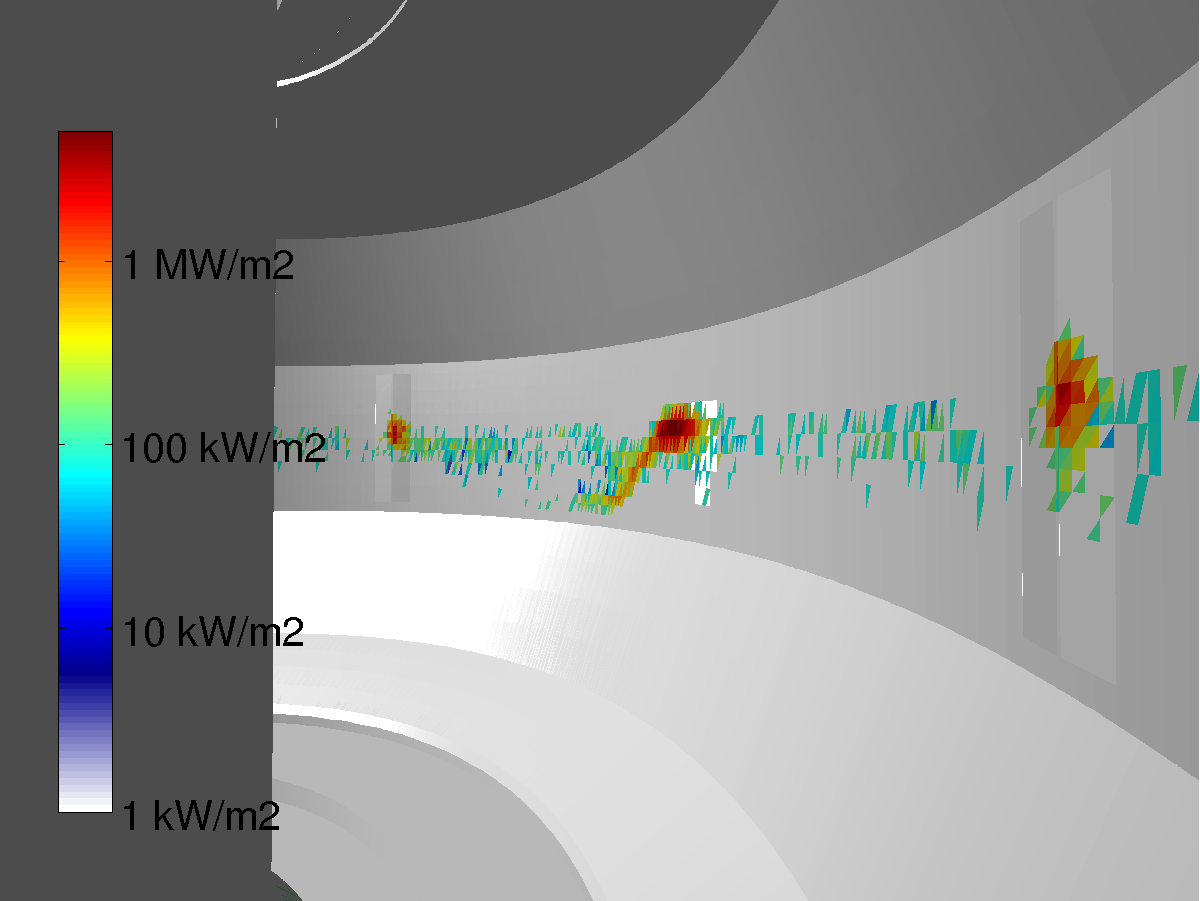}
\caption{\label{fig:d3d-wall} The simulated power load due to lost NBI ions on the first wall of DIII-D in the test blanket module mock-up experiment. The TBM mock-up module is located inside the wall in the middle of the figure.}
\end{figure}

\begin{figure}
\centering
\includegraphics[width=.45\linewidth]{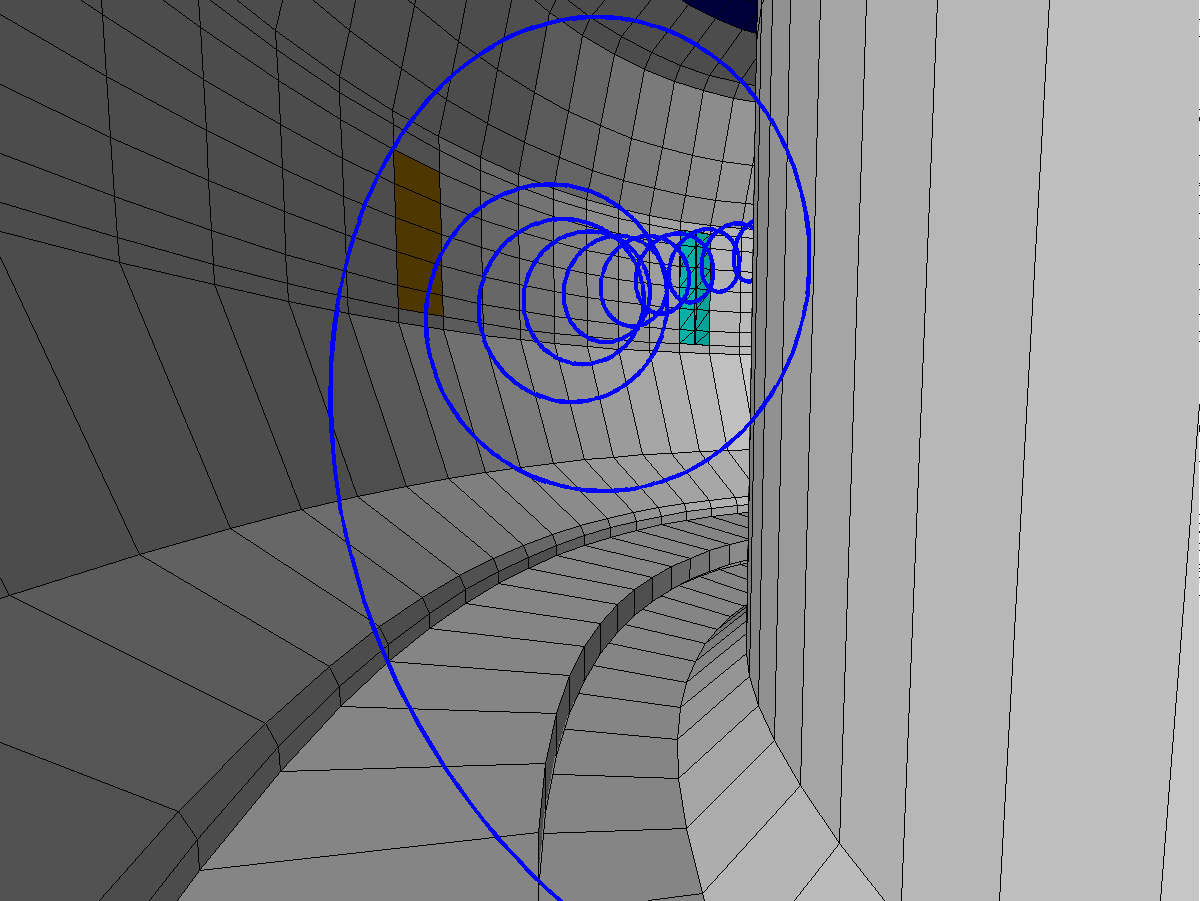}
\includegraphics[width=.45\linewidth]{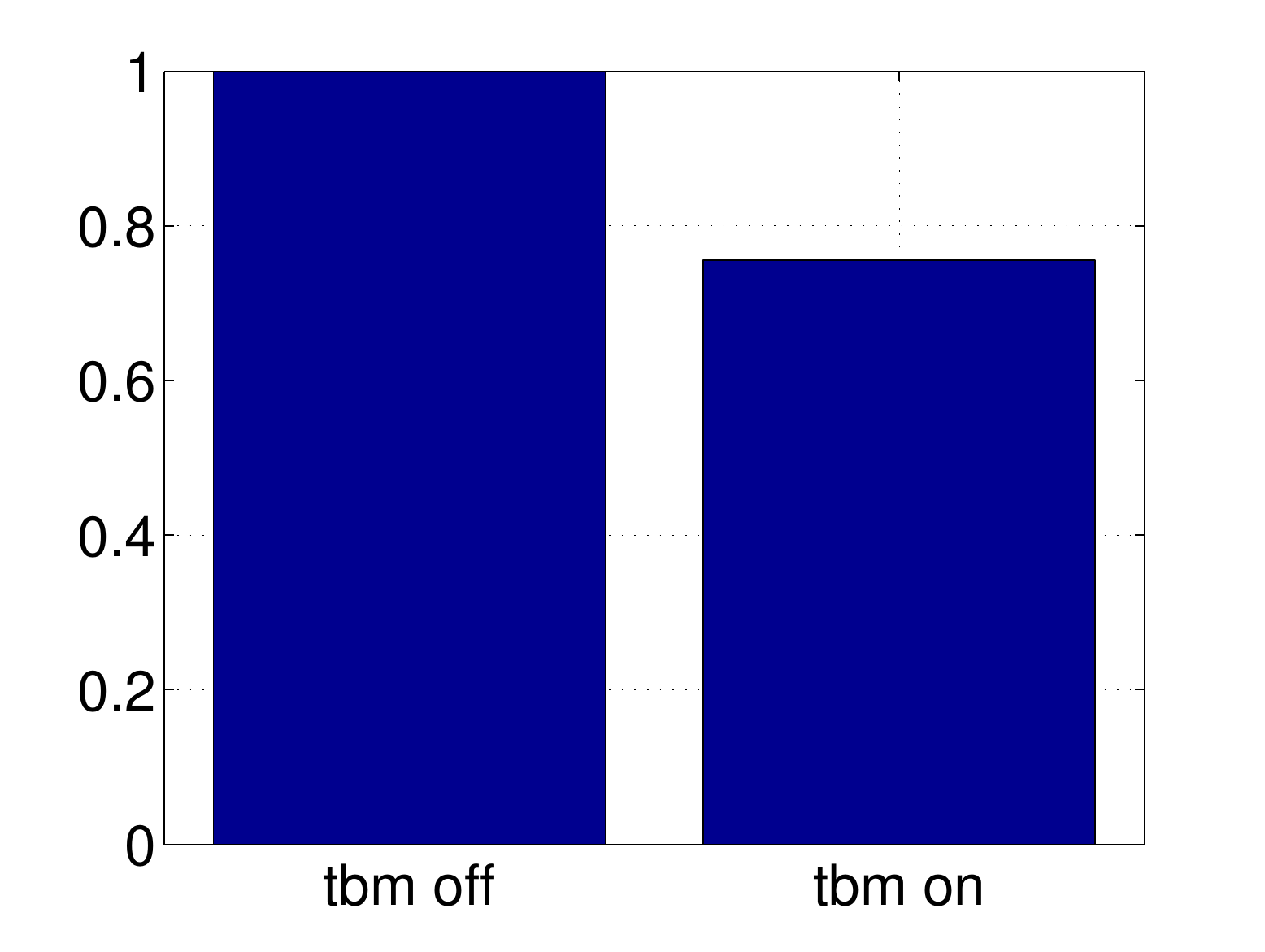}
\caption{\label{fig.d3d-T} Simulation of 1 MeV DD-fusion tritons in DIII-D. (a) The full gyro-orbit of a 1 MeV tritium ion. (b) The number of confined tritons without and with the TBM mock-up coils, normalized to the number without the mock-up.}
\end{figure}

\subsection{Tracer injection experiments at ASDEX Upgrade}
Besides wall damage due to fast ions, another potential show stopper for ITER operation is provided by impurities in the form of tritium compounds. ITER is a nuclear device and, therefore, licensing sets strict limits to the amount of volatile tritium contained in the vessel. Therefore erosion, transport and deposition of impurities likely to form tritium compounds has to be understood, so that the deposition regions can be predicted and, if needed, occasionally cleaned.

ASDEX Upgrade has invested over ten years on studies of impurity migration in the form of tracer injection with mainly carbon-13 serving as the tracer, see, e.g., for more details~\citep{hakola13b}. The experiment is always carried out on the last day of an experimental campaign, and a large number of identical plasma shots are carried out with the injection. This is to provide constant plasma conditions for the migrating impurities which is beneficial for modelling. After the experiment, selected wall tiles are removed for surface analysis. 

In an experiment carried out in 2007, the post mortem analysis brought the puzzling result that over 90 \% of the injected $^{13}$C was missing when assuming toroidally symmetric deposition~\citep{hakola10}. No obvious experimental explanation for the disappearance of the carbon was found.

At this point it was decided that it was worth studying the problem with ASCOT. This required some refurbishing of the code, most notably including relevant atomic reactions and background plasma flows that had been found to play a very important role for impurity transport in the scrape-off layer. Typically impurity transport simulations assume an axisymmetric wall geometry and magnetic field, whereas with ASCOT both the toroidal ripple and realistic 3D wall structure could be included, see Fig.~\ref{fig:C-13}.
\begin{figure}
\centering
\includegraphics[width=0.7\textwidth]{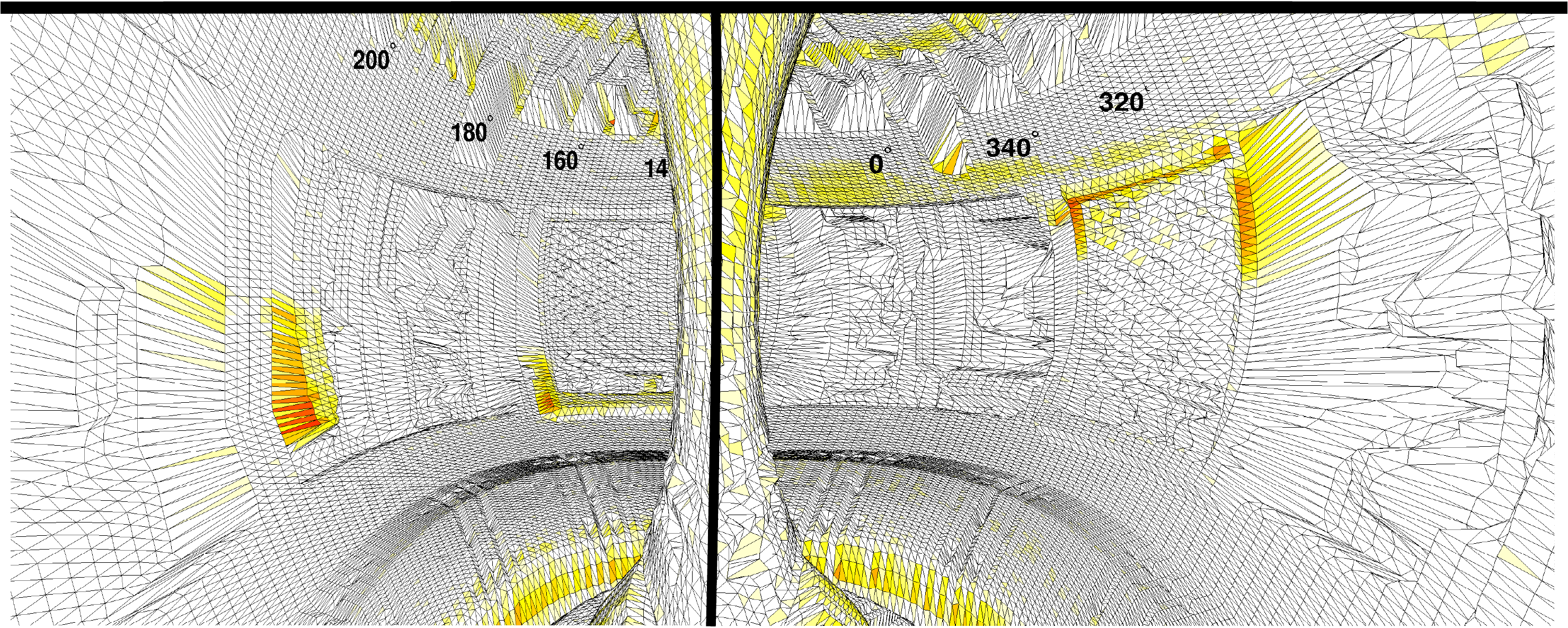}
\caption{\label{fig:C-13} Three-dimensional representation of $^{13}$C deposition in ASDEX Upgrade as simulated with ASCOT. Highly localized deposition can be observed on the protruding wall structures such as limiters on the outer wall. }
\end{figure}

According to the simulation results shown in~\citep{miettunen:NF2012}, the deposition pattern on the outer wall is far from uniform but, rather, is concentrated on protruding elements such as antenna limiters as indicated in the figure. These results provided at least a partial explanation to the mystery of the missing $^{13}$C since experimentally outer wall deposition had been studied only from a region receiving rather low deposition according to the simulations. Therefore, the ASCOT simulations indicated the assumption of toroidally symmetric impurity deposition to be invalid on the outer wall which was then also experimentally confirmed in~\citep{hakola13a}.

\section{Summary and outlook}
This contribution reviewed the Fokker-Planck theory of minority particle populations in plasmas in both particle and guiding-center phase-space. Along the way, also the Lie-transformation, needed for constructing the guiding-center formalism, was outlined. The emphasis was in solving the distribution function of \emph{minority ions}, such as energetic particles or impurity ions in tokamak plasmas. The proposed solution method is Monte Carlo because it requires the least amount of simplifying approximations. This approach gives birth to the \emph{Monte Carlo -based orbit-following codes}. 

After the formal part, the attention was given to high-performance computing. First, the need for super computers in solving the Fokker-Planck equation using Monte Carlo method was justified. Then an introduction to the structure and functioning of a supercomputer was given, using Monte Carlo -based orbit-following codes as an example. 

The contribution was finished with examples where the approach outlined here had been applied. The examples were obtained using the Finnish Monte Carlo -based orbit-following code ASCOT, that was briefly introduced before the examples.

The future of Monte Carlo -based orbit-following codes appears bright: it is fairly straightforward to include more physics to it via different interactions with the background. However, this has to be done with care so that the structure of the formalism is not broken and no numerical physics is introduced to the system. %

\appendix
\section{}
\label{app:explicit_lie_transformation}
As the goal of the guiding-center transformation is to rid the particle phase-space Lagrangian one-form from the gyromotion, we need to know what a Lie-derivative actually does for a differential one-form. As defined in~\citep{ralph1978foundations}, a Lie-derivative of a differential one-form $\gamma=\gamma_{\alpha}dz^{\alpha}$ is another differential one-form
\begin{align}
\mathcal{L}_{G}\gamma=G^{\alpha}\omega_{\alpha\beta}dz^{\beta}+\partial_{\beta}\left(G^{\alpha}\gamma_{\alpha}\right)dz^{\beta},
\end{align}
where the second term on the right-hand-side is a total differential $\partial_{\beta}\left(G^{\alpha}\gamma_{\alpha}\right)dz^{\beta}=d\left(G^{\alpha}\gamma_{\alpha}\right)$, and $\omega_{\alpha\beta}=\partial_{\alpha}\gamma_{\beta}-\partial_{\beta}\gamma_{\alpha}$. From classical mechanics, one may recall that adding a total derivative to the Lagrangian does not change the equations of motion, so we focus on the first term $G^{\alpha}\omega_{\alpha\beta}dz^{\beta}$ and how that can be used to reach our goals. Now, consider the Lie-transformation, defined as the exponential serie operator%
\begin{align}
\exp\left(\mathcal{L}_G\right)\gamma_{\alpha}dz^{\alpha}=\left(\gamma_{\beta}+G^{\alpha}\omega_{\alpha\beta}\right)dz^{\beta}+d\left(G^{\alpha}\gamma_{\alpha}\right)+\mathrm{''other\ terms''},
\end{align}
where the ''other terms'' include the rest of the exponential serie, i.e., 
\begin{align}
\mathrm{''other\ terms''} = \frac{1}{2!}\mathcal{L}_G\mathcal{L}_G\gamma+\frac{1}{3!}\mathcal{L}_G\mathcal{L}_G\mathcal{L}_G\gamma+\dots
\end{align} 
If some of the $\gamma_{\alpha}$ would depend on unwanted coordinates, e.g., the gyro-angle, we could formally eliminate that by choosing the components $G^{\alpha}$ wisely, if the ''other terms'' would be neglected. In principle, one could for example solve for matrix equation $\omega_{\alpha\beta}G^{\beta}+\gamma_{\alpha}=0$ to set the result of the transformation to zero. The generating function is thus solved from the condition that the components of the transformed one-form become independent of unwanted coordinates. Because of the ''other terms'', a single transformation is not enough to completely eliminate the gyroangle from the components of the transformed fundamental one-form and from the Hamiltonian. Even if the vector field $G$ is chosen to eliminate the gyro-angle from the first term, so that $\left(\gamma_{\beta}+G^{\alpha}\omega_{\alpha\beta}\right)dz^{\beta}$ becomes independent of it, the other terms will depend on the gyro-angle through the vector field $G$. Thus, more free parameters are needed.

In the Lie-transform perturbation theory, a sequence of individual transformations, $\mathcal{T}_n=\exp\left(\epsilon^n\mathcal{L}_{G_n}\right)$, is used instead, with each transformation ordered using a dimensionless \emph{ordering parameter} $\epsilon$. The guiding-center transformation is defined as an asymptotic serie
\begin{equation}
\mathcal{T}_{gc}^{\pm} = \exp\left(\pm\sum_{n=1}^{\infty}\epsilon^n\mathcal{L}_{G_n}\right),
\end{equation}
which up to second order is explicitly given by the sequences
\begin{align}
\mathcal{T}_{gc} =& 1 + \epsilon \mathcal{L}_{G_1} +  \epsilon^2(\mathcal{L}_{G_2} + \frac{1}{2} \mathcal{L}_{G_1}^2)+\dots,\\
\mathcal{T}_{gc}^{-1} =&1 - \epsilon \mathcal{L}_{G_1} -  \epsilon^2(\mathcal{L}_{G_2} - \frac{1}{2} \mathcal{L}_{G_1}^2)+\dots.
\end{align}
Now the ''other terms'' can be taken care of, order by order, by defining always a new generating vector field: in first order, choose $G_1$, in second order choose $G_2$ according to the choice of $G_1$, and so on.}

\bibliographystyle{jpp}
\bibliography{IIS2014_TKS_final}

\begin{thebibliography}{27}
\expandafter\ifx\csname natexlab\endcsname\relax\def\natexlab#1{#1}\fi

\bibitem[Abraham \& Marsden(1978)]{ralph1978foundations}
{\sc Abraham, Ralph \& Marsden, Jerrold~E.} 1978 {\em Foundations of
  Mechanics\/}. AMS Chelsea Pub./American Mathematical Society.

\bibitem[Alfv\'{e}n(1940)]{alfven1940}
{\sc Alfv\'{e}n, H.} 1940 On the motion of a charged particle in a magnetic
  field. {\em Arkiv Mat. Astr. Fysik\/} .

\bibitem[Arnold(1989)]{arnold1989}
{\sc Arnold, V.~I.} 1989 {\em Mathematical Methods of Classical Mechanics, 2nd.
  ed.\/}. Springer-Verlag Pub. Co.

\bibitem[Brizard(1995)]{brizard:guidingCenterTransformationPOP1995}
{\sc Brizard, A.J.} 1995 Nonlinear gyrokinetic vlasov equation for toroidally
  rotating axisymmetric tokamaks. {\em Physics of Plasmas\/} {\bf 2}~(2),
  459--471.

\bibitem[Brizard(2004)]{brizard:fokkerPlanck}
{\sc Brizard, Alain~J.} 2004 A guiding-center fokker--planck collision operator
  for nonuniform magnetic fields. {\em Physics of Plasmas\/} {\bf 11}~(9),
  4429--4438.

\bibitem[Cary \& Brizard(2009)]{RevModPhys.81.693}
{\sc Cary, J.R. \& Brizard, A.J.} 2009 Hamiltonian theory of guiding-center
  motion. {\em Rev. Mod. Phys.\/} {\bf 81}, 693--738.

\bibitem[Decker {\em et~al.\/}(2010)Decker, Peysson, Brizard \&
  Duthoit]{decker:orbitAveragedFokkerPlanck}
{\sc Decker, J., Peysson, Y., Brizard, A.~J. \& Duthoit, F.-X.} 2010
  Orbit-averaged guiding-center fokker--planck operator for numerical
  applications. {\em Physics of Plasmas\/} {\bf 17}~(11), 112513.

\bibitem[Flanders(1989)]{flanders1989differential}
{\sc Flanders, H.} 1989 {\em Differential Forms with Applications to the
  Physical Sciences\/}. Dover Publications.

\bibitem[Hakola {\em et~al.\/}(2013{\natexlab{{\em a\/}}})Hakola, Airila,
  Bj\"orkas, Borodin, Brezinsek, Coad, Groth, J\"arvinen, Kirschner,
  Koivuranta, Krieger, Kurki-Suonio, Likonen, Lindholm, Makkonen, Mayer,
  Miettunen, M\"uller, Neu, Petersson, Rohde, Rubel, Widdowson, {the ASDEX
  Upgrade Team} \& {JET-EFDA Contributors}]{hakola13b}
{\sc Hakola, A, Airila, M~I, Bj\"orkas, C, Borodin, D, Brezinsek, S, Coad, J~P,
  Groth, M, J\"arvinen, A, Kirschner, A, Koivuranta, S, Krieger, K,
  Kurki-Suonio, T, Likonen, J, Lindholm, V, Makkonen, T, Mayer, M, Miettunen,
  J, M\"uller, H~W, Neu, R, Petersson, P, Rohde, V, Rubel, M, Widdowson, A,
  {the ASDEX Upgrade Team} \& {JET-EFDA Contributors}} 2013{\natexlab{{\em
  a\/}}} {Global migration of impurities in tokamaks}. {\em Plasma Physics and
  Controlled Fusion\/} {\bf 55}, 124029.

\bibitem[Hakola {\em et~al.\/}(2013{\natexlab{{\em b\/}}})Hakola, Koivuranta,
  Likonen, Groth, Kurki-Suonio, Lindholm, Miettunen, Krieger, Mayer,
  M\"{u}ller, Neu, Rohde, Petersson \& {the ASDEX Upgrade Team}]{hakola13a}
{\sc Hakola, A., Koivuranta, S., Likonen, J., Groth, M., Kurki-Suonio, T.,
  Lindholm, V., Miettunen, J., Krieger, K., Mayer, M., M\"{u}ller, H.W., Neu,
  R., Rohde, V., Petersson, P. \& {the ASDEX Upgrade Team}} 2013{\natexlab{{\em
  b\/}}} {Global migration of $^{13}$C impurities in high-density L-mode
  plasmas in ASDEX Upgrade}. {\em Journal of Nuclear Materials\/} {\bf
  438}~(0), S694--S697, proceedings of the 20th International Conference on
  Plasma-Surface Interactions in Controlled Fusion Devices.

\bibitem[Hakola {\em et~al.\/}(2010)Hakola, Likonen, Aho-Mantila, Groth,
  Koivuranta, Krieger, Kurki-Suonio, Makkonen, Mayer, M\"uller, Neu, Rohde \&
  {the ASDEX Upgrade Team}]{hakola10}
{\sc Hakola, A, Likonen, J, Aho-Mantila, L, Groth, M, Koivuranta, S, Krieger,
  K, Kurki-Suonio, T, Makkonen, T, Mayer, M, M\"uller, H~W, Neu, R, Rohde, V \&
  {the ASDEX Upgrade Team}} 2010 {Migration and deposition of $^{13}$C in the
  full-tungsten ASDEX Upgrade tokamak}. {\em Plasma Physics and Controlled
  Fusion\/} {\bf 52}, 065006.

\bibitem[Helander \& Sigmar(2005)]{helander2005}
{\sc Helander, Per \& Sigmar, Dieter~J} 2005 Collisional transport in
  magnetized plasmas. {\em Collisional Transport in Magnetized Plasmas, by Per
  Helander, Dieter J. Sigmar, Cambridge, UK: Cambridge University Press,
  2005\/} {\bf 1}.

\bibitem[Hirvijoki(2014)]{EeroPHD}
{\sc Hirvijoki, Eero} 2014 Theory and models for monte carlo simulations of
  minority particle populations in tokamak plasmas. PhD thesis, Aalto
  University, School of Science.

\bibitem[Hirvijoki {\em et~al.\/}(2014)Hirvijoki, Asunta, Koskela,
  Kurki-Suonio, Miettunen, Sipil{\"a}, Snicker \&
  {\"A}k{\"a}slompolo]{ascot4ref}
{\sc Hirvijoki, E., Asunta, O., Koskela, T., Kurki-Suonio, T., Miettunen, J.,
  Sipil{\"a}, S., Snicker, A. \& {\"A}k{\"a}slompolo, S.} 2014 {ASCOT}: Solving
  the kinetic equation of minority particle species in tokamak plasmas. {\em
  Computer Physics Communications\/} {\bf 185}~(4), 1310--1321.

\bibitem[Hirvijoki {\em et~al.\/}(2013)Hirvijoki, Brizard, Snicker \&
  Kurki-Suonio]{hirvijoki:092505:2013:pop}
{\sc Hirvijoki, E., Brizard, A., Snicker, A. \& Kurki-Suonio, T.} 2013 {Monte
  Carlo} implementation of a guiding-center fokker-planck kinetic equation.
  {\em Physics of Plasmas\/} {\bf 20}~(9), 092505.

\bibitem[Hirvijoki {\em et~al.\/}(2012)Hirvijoki, Snicker, Korpilo, Lauber,
  Poli, Schneller \& Kurki-Suonio]{mhdpapru}
{\sc Hirvijoki, E., Snicker, A., Korpilo, T., Lauber, P., Poli, E., Schneller,
  M. \& Kurki-Suonio, T.} 2012 Alfv\'{e}n {Eigenmodes} and {Neoclassical}
  tearing modes for orbit-following implementations. {\em Computer Physics
  Communications\/} {\bf 183}~(12), 2589--2593.

\bibitem[Ichimaru(1973)]{ichimaru1973basic}
{\sc Ichimaru, S.} 1973 {\em Basic Principles of Plasma Physics: A Statistical
  Approach\/}. W. A. Benjamin.

\bibitem[Kramer {\em et~al.\/}(2011)Kramer, Budny, Ellis, Gorelenkova,
  Heidbrink, Kurki-Suonio, Nazikian, Salmi, Schaffer, Shinohara, Snipes, Spong,
  Koskela \& Zeeland]{kramer:nf:51:103029}
{\sc Kramer, G.J., Budny, B.V., Ellis, R., Gorelenkova, M., Heidbrink, W.W.,
  Kurki-Suonio, T., Nazikian, R., Salmi, A., Schaffer, M.J., Shinohara, K.,
  Snipes, J.A., Spong, D.A., Koskela, T. \& Zeeland, M.A.~Van} 2011 {Fast-ion
  effects during test blanket module simulation experiments in DIII-D}. {\em
  Nuclear Fusion\/} {\bf 51}~(10), 103029.

\bibitem[Kramer {\em et~al.\/}(2013)Kramer, McLean, Brooks, Budny, Chen,
  Heidbrink, Kurki-Suonio, Nazikian, Koskela, Schaffer, Shinohara, Snipes \&
  Zeeland]{Kramer13_fast_ion_heat_loads_TBM}
{\sc Kramer, G.J., McLean, A., Brooks, N., Budny, R.V., Chen, X., Heidbrink,
  W.W., Kurki-Suonio, T., Nazikian, R., Koskela, T., Schaffer, M.J., Shinohara,
  K., Snipes, J.A. \& Zeeland, M.A.~Van} 2013 Simulation of localized fast-ion
  heat loads in test blanket module simulation experiments on diii-d. {\em
  Nuclear Fusion\/} {\bf 53}~(12), 123018.

\bibitem[Littlejohn(1979)]{littlejohn:guidingCenterHamiltomianJMP1979}
{\sc Littlejohn, R.G.} 1979 {A guiding center Hamiltonian: A new approach}.
  {\em Journal of Mathematical Physics\/} {\bf 20}~(12), 2445--2458.

\bibitem[Littlejohn(1982)]{littlejohn:hamiltonianPerturbationTheoryJMP1982}
{\sc Littlejohn, R.G.} 1982 Hamiltonian perturbation theory in noncanonical
  coordinates. {\em Journal of Mathematical Physics\/} {\bf 23}~(5), 742--747.

\bibitem[Littlejohn(1983)]{littlejohn:variationalPrinciplesJPP1983}
{\sc Littlejohn, R.G.} 1983 Variational principles of guiding centre motion.
  {\em Journal of Plasma Physics\/} {\bf 29}, 111--125.

\bibitem[Miettunen {\em et~al.\/}(2012)Miettunen, Kurki-Suonio, Makkonen,
  Groth, Hakola, Hirvijoki, Krieger, Likonen, {\"A}k{\"a}slompolo \& {the ASDEX
  Upgrade Team}]{miettunen:NF2012}
{\sc Miettunen, J., Kurki-Suonio, T., Makkonen, T., Groth, M., Hakola, A.,
  Hirvijoki, E., Krieger, K., Likonen, J., {\"A}k{\"a}slompolo, S. \& {the
  ASDEX Upgrade Team}} 2012 The effect of non-axisymmetric wall geometry on
  $^{13}${C} transport in {ASDEX Upgrade}. {\em Nuclear Fusion\/} {\bf 52}~(3),
  032001.

\bibitem[Northrop(1961)]{gca1961}
{\sc Northrop, T.G.} 1961 The guiding center approximation to charged particle
  motion. {\em Annals of Physics\/} {\bf 15}~(1), 79--101.

\bibitem[Northrop(1963)]{northrop:adiabaticMotionReview1963}
{\sc Northrop, T.G.} 1963 Adiabatic charged-particle motion. {\em Reviews of
  Geophysics\/} {\bf 1}~(3), 283--304.

\bibitem[{\O}ksendal(2003)]{oksendal2003stochastic}
{\sc {\O}ksendal, B.} 2003 {\em Stochastic Differential Equations: An
  Introduction with Applications\/}. Springer.

\bibitem[Rosenbluth {\em et~al.\/}(1957)Rosenbluth, MacDonald \&
  Judd]{PhysRev.107.1:rosenbluth}
{\sc Rosenbluth, Marshall~N., MacDonald, William~M. \& Judd, David~L.} 1957
  Fokker-planck equation for an inverse-square force. {\em Phys. Rev.\/} {\bf
  107}, 1--6.

\end{thebibliography}
\end{document}